\documentclass[aps,twocolumn,superscriptaddress]{revtex4-1} 
\usepackage{graphicx}
\usepackage{xcolor}
\usepackage{subfigure}
\usepackage{mathtools}
\usepackage{physics,amsmath}
\usepackage{epsfig}
\usepackage{chemfig}
\usepackage[colorlinks= true,urlcolor=blue,linkcolor= blue,citecolor=blue,bookmarks=false,pdfstartview=]{hyperref}
\usepackage{siunitx}
\usepackage{bm} 
\usepackage{xr}
\usepackage{xcolor}
\usepackage{placeins}
\usepackage{braket}
\usepackage{epsfig}
\usepackage{breqn}
\usepackage{etoolbox}
\usepackage{multirow}
\makeatletter
\preto\maketitle{%
  \begingroup\lccode`~=`,
  \lowercase{\endgroup
  \let\saved@breqn@active@comma~
  \let~}\active@comma 
}
\appto\maketitle{%
  \begingroup\lccode`~=`,
  \lowercase{\endgroup
  \let~}\saved@breqn@active@comma 
}
\makeatother

\begin{document}

\title{Pseudo-easy-axis anisotropy in antiferromagnetic $S=1$ diamond-lattice systems}

\author{S. Vaidya}
\email{s.vaidya@warwick.ac.uk}
\affiliation{Department of Physics, University of Warwick, Gibbet Hill Road, Coventry, CV4 7AL, UK}
\author{A. Hern\'{a}ndez-Meli\'{a}n}
\affiliation{Department of Physics, Durham University, South Road, Durham, DH1 3LE, United Kingdom}
\author{J. P. Tidey}
\affiliation{Department of Physics, University of Warwick, Gibbet Hill Road, Coventry, CV4 7AL, UK}
\author{S. P. M. Curley}
\author{S. Sharma}
\affiliation{Department of Physics, University of Warwick, Gibbet Hill Road, Coventry, CV4 7AL, UK}
\author{P. Manuel}
\affiliation{ISIS Pulsed Neutron Source, STFC Rutherford Appleton Laboratory, Didcot, Oxfordshire OX11 0QX, United Kingdom}
\author{C. Wang}
\affiliation{Laboratory for Muon Spin Spectroscopy, Paul Scherrer Institute, 5232 Villigen, Switzerland}
\author{G. L. Hannaford}
\affiliation{Department of Physics, Durham University, South Road, Durham, DH1 3LE, United Kingdom}
\author{S. J. Blundell}
\affiliation{Department of Physics, Clarendon Laboratory, University of Oxford, Parks Road, Oxford, OX1 3PU, United Kingdom}
\author{Z. E. Manson}
\author{J. L. Manson}\thanks{Deceased 7 June 2023.}
\affiliation{Department of Chemistry and Biochemistry, Eastern Washington University, Cheney, Washington 99004, USA}
\author{J. Singleton}
\affiliation{National High Magnetic Field Laboratory (NHMFL), Los Alamos National Laboratory, Los Alamos, New Mexico 87545, USA}
\author{T. Lancaster}
\affiliation{Department of Physics, Durham University, South Road, Durham, DH1 3LE, United Kingdom}
\author{R. D. Johnson}
\affiliation{Department of Physics and Astronomy, University College London, London, UK}
\affiliation{London Centre for Nanotechnology and Department of Physics and Astronomy, University College London, London WC1E 6BT, UK}
\author{P. A. Goddard}
\email{p.goddard@warwick.ac.uk}
\affiliation{Department of Physics, University of Warwick, Gibbet Hill Road, Coventry, CV4 7AL, UK}

\begin{abstract}
We investigate the magnetic properties of $S=1$ antiferromagnetic diamond-lattice, Ni$X_{2}$(pyrimidine)$_{2}$ ($X$\,=\,Cl, Br), hosting a single-ion anisotropy (SIA) orientation which alternates between neighbouring sites. Through neutron diffraction measurements of the $X$\,=\,Cl compound, the ordered state spins are found to align collinearly along a pseudo-easy-axis, a unique direction created by the intersection of two easy planes. Similarities in the magnetization, exhibiting spin-flop transitions, and the magnetic susceptibility in the two compounds imply that the same magnetic structure and a pseudo-easy-axis is also present for $X$\,=\,Br. We estimate the Hamiltonian parameters by combining analytical calculations and Monte-Carlo (MC) simulations of the spin-flop and saturation field. The MC simulations also reveal that the spin-flop transition occurs when the applied field is parallel to the pseudo-easy-axis. Contrary to conventional easy-axis systems, there exist field directions perpendicular to the pseudo-easy-axis for which the magnetic saturation is approached asymptotically and no symmetry-breaking phase transition is observed at finite fields.
\end{abstract}

\maketitle
\section{Introduction}
A fundamental goal in condensed matter physics has been the control of dominant energy scales in magnetic systems, allowing experimental exploration of various emergent phases and quantum phenomena. To this end, molecule-based magnets, where molecular ligands link magnetic metal ions, have proven particularly effective~\cite{Dender_1997, Hong_2006, Zapf_2006, Brambleby_2017, Willaims_2020_haldane}. The wide choices of ligands available provide a high level of control over the final crystal structure and thus control over various aspects of the magnetic properties~\cite{Thorarinsdottir_MOM_rev_2020, Coronado_2020_molecular, Chilton_2022_molecular}. For example, a mix of bridging and non-bridging ligands can be used to create quasi-low dimensional magnetic structures ~\cite{Goddard_dimensional_2012, Pitcarin_2023, Wang_dimensional_2024}. The control over the magnetic properties has been achieved through various means, including the substitution of ligands~\cite{Schlueter_2012, Cortijo_2013, Cortijo_2014, Liu_2016_correlations, Kubus_2018, Blackmore-correlations}, metal-ions~\cite{Pitcarin_2023, Geers_2023_NCS} and the counter-ions~\cite{Woodward_counter-ion_2007, Manson_counter-ion_2021}, and by applying hydrostatic pressure~\cite{Pajerowski_2022, Coak_2023, Geers_preesure_2023, Povarov_2024}. Moreover, organic ligands such as pyrimidine (pym = C$_{4}$H$_{4}$N$_{2}$) provide the flexibility to create structurally non-trivial systems such as staggered and chiral antiferromagnetic (AFM) $S=1/2$ spin chains~\cite{Feyerherm_stag_2000, Zvyagin_SG_2004, Liu_chiral_2019, Huddart_spin_transport_2021}. In these systems, the alternating orientation of the spin octahedra between nearest neighbours leads to magnetic properties that differ significantly from conventional spin-half chains. Such properties were explained by the application of the sine-Gordon model of quantum field theory~\cite{Oshikawa_SG_1997, Affleck_SG_1999}. 

In the present study, we explore the effects of halide substitution and alternating local octahedra orientation in the context of $S=1$ three-dimensional diamond-lattice systems Ni$X_{2}$(pym)$_{2}$, where $X$\,=\,Cl, Br. The zero-field-splitting in these systems presents the possibility of single-ion-anisotropy (SIA), with energy $D$, which follows the local spin octahedra and alternates in orientation on neighbouring spin sites. Previous studies on the $S=2$ isostructural system FeCl$_{2}$(pym)$_{2}$, showed that an alternating easy-axis SIA direction results in a large canting angle of $14^{\circ}$ and weak-ferromagnetism significantly larger than that which is conventionally caused by the Dzyalonzinskii-Moriya (DM) interaction alone~\cite{FeCl2}. A subsequent study on NiCl$_{2}$(pym)$_{2}$, under applied hydrostatic pressure, observed temperature-dependent magnetic susceptibility, $\chi(T)$, curves that exhibited similarities to the $\chi(T)$ of FeCl$_{2}$(pym)$_{2}$. This led to the assumption that the Ni system hosted a similar spin-canted ground state~\cite{XCl2_pressure}. However, neutron diffraction or magnetic hysteresis measurement, showing a zero-field remanent magnetization, were not available to support this claim. The application of pressure was reported to increase the magnetic ordering temperature, $T_{\text{N}}$, due to increasing orbital overlap in the Ni---N (N from the pym) bond. Quantitative modelling of the pressure-dependent $T_{N}$ found that $J\approx7$\,K but an estimation for the size of the SIA was not possible. More recently, studies on a family of metal-organic magnets $M$Cl$_{2}L$, where $M$ = Ni and Fe, $L$ = (pym) and 2,1,3-benzothiadiazole (btd) also find that an alternating easy-axis induces a canted magnetic structure~\cite{XCl2L_Jem_2024}.

In contrast to these studies, our neutron diffraction measurement reveals a collinear magnetic structure in NiCl$_{2}$(pym)$_{2}$, which is attributed to an \emph{easy-plane} SIA that alternates in orientation. The intersection of these alternating planes coupled with the nearest neighbour antiferromagnetic exchange interaction creates a unique pseudo-easy-axis along which the moments align. Given the similarities in the crystal structures (determined using electron diffraction), magnetometry and muon-spin rotations ($\mu^{+}$SR) data, we propose that a pseudo-easy-axis is also present in the $X$\,=\,Br case. The magnetic order present in these systems is consistent with the negligible zero-field remanent magnetization in measurements of the field-dependent magnetization, $M(H)$. Additionally, we find that the pseudo-easy-axis, similar to a real easy-axis, results in a spin-flop transition seen in the $M(H)$ data. Monte-Carlo (MC) simulations of $M(H)$ and mean-field calculations of the spin-flop and saturation field, $H_{\text{SF}}$ and $H_{\text{C1}}$ respectively, are used to determine $J$ and $D$ for $X$\,=\,Cl and Br. For $X$\,=\,Br, we find that $D$ is reduced while a secondary AFM interaction mediated by the $X\cdot\cdot\cdot X$ exchange bonds is increased. This is in keeping with the trends established in the magnetic chain compounds Ni$X_{2}$(3,5-lut)$_{4}$  ($X$\,=\,HF2, Cl, Br, or I)~\cite{Blackmore-correlations} and Ni$X_{2}$(pyz)$_{2}$ (pyz = pyrazine C$_{4}$H$_{4}$N$_{2}$; $X$\,=\,Cl, Br, I, NCS)~\cite{Liu_2016_correlations}. Finally, MC simulations of single-crystal $M(H)$ curves reveal that a magnetic saturation phase transition only occurs for fields applied along certain directions.


\section{Results and Discussion}
\subsection{Crystal structure}
\label{sec: struct}
\begin{figure}
    \centering
    \includegraphics[width= \linewidth]{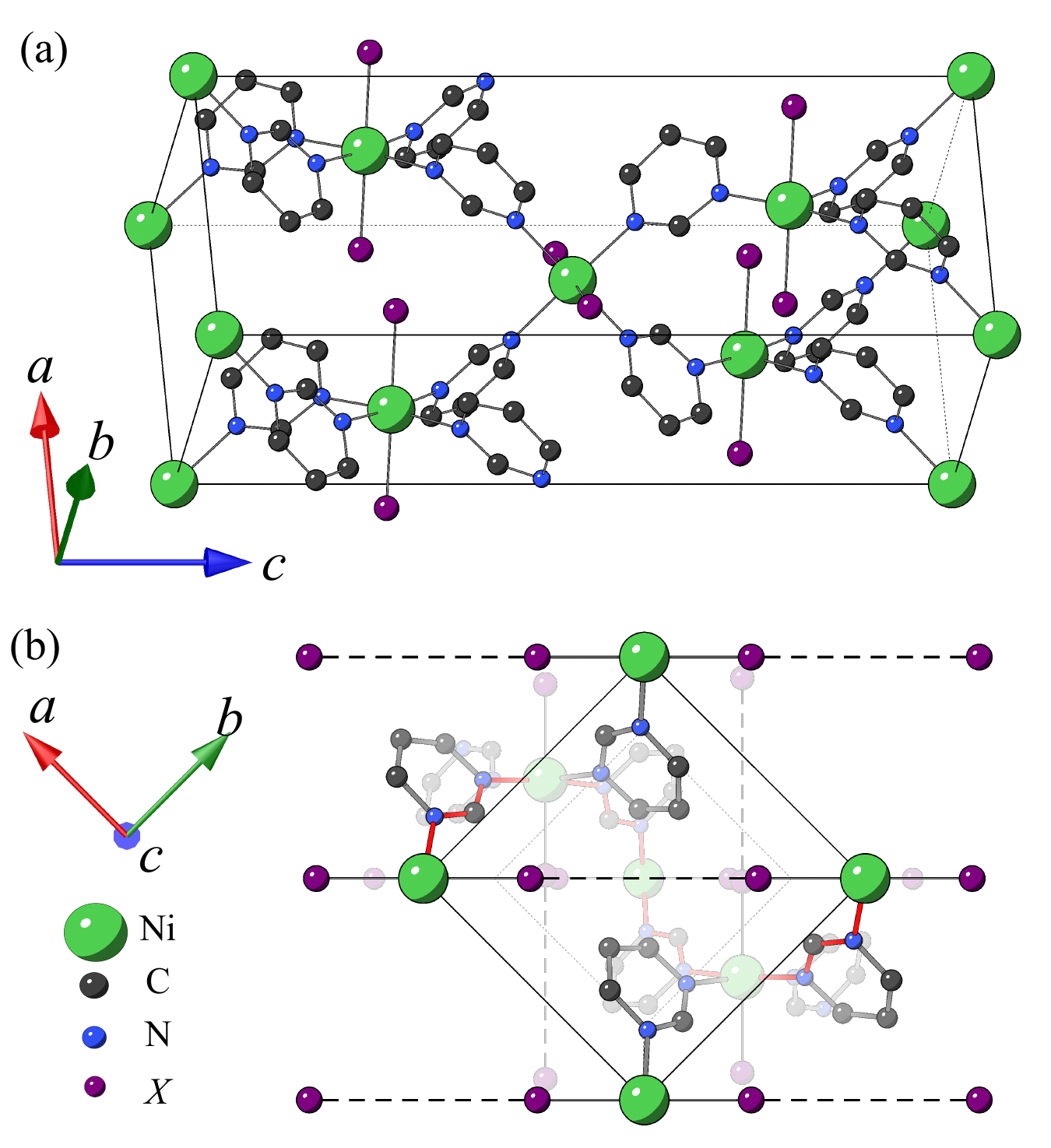}
    \caption{(a) The $I4_{1}22$ crystal structure of Ni$X_{2}$(pym)$_{2}$ where $X$\,=\,Cl or Br and pym\,=\,pyrimidine. (b) Displays the view down the $c$-axis. In this structure, the pym ligands bridge neighbouring Ni ions and mediate the primary exchange interaction, forming a three-dimensional diamond network. Between neighbouring Ni ions connected by pym, the Ni---$X$ axial bond direction rotates by $90^{\circ}$. The black-dashed lines highlight the weak $X\cdot\cdot\cdot X$ exchange bonds that mediate the secondary exchange interaction along the Ni---$X\cdot\cdot\cdot X$---Ni chains. The red bond highlights the exchange pathway, in the Ni---pym network, between two Ni(II) ions connected by the $X\cdot\cdot\cdot X$ exchange bond.}
    \label{fig: NiX2_struct}
\end{figure}

The crystal structures of both $X$\,=\,Cl and Br compound are solved using electron diffraction measurements at $T=120$\,K with lattice parameters obtained from powder from powder X-ray diffraction measurements at the same temperature. The details of these measurements and the structural refinement can be found in the supplemental material~\cite{supplementary}. The resulting crystal structure of Ni$X_{2}$(pym)$_{2}$ is illustrated in Fig.~\ref{fig: NiX2_struct} and Table~\ref{tab: lattic_params} provides the lattice parameters.

\begin{table}
\caption{\label{tab: lattic_params}%
Lattice parameters for the $I4_{1}22$ structure of Ni$X_{2}$(pym)$_{2}$, X=Cl and X=Br, determined using powder X-ray diffraction at $T=120$\,K. 
}
\begin{ruledtabular}
\begin{tabular}{ccc}
\textrm{}&
\textrm{$a, b$\,($\si{\angstrom}$)}&
\textrm{$c$\,($\si{\angstrom}$)}\\
\colrule
NiCl$_{2}$(pym)$_{2}$ & $7.3576(2)$ & $19.4956(5)$\\
NiBr$_{2}$(pym)$_{2}$ & $7.50690(6)$ & $19.5378(2)$ \\
\end{tabular}
\end{ruledtabular}
\end{table}

The $X$\,=\,Cl and $X$\,=\,Br materials are isostructural and both crystallise in the $I4_{1}22$ space group. These materials consist of three-dimensional diamond-like networks of Ni(II) ions connected by the pym ligand. The Ni(II) ions sit at the centre of axially distorted octahedra consisting of four equatorial bonds to the N atoms of the pym rings and two longer axial coordination bonds to the halides.

Within the unit cell Ni(II) ions reside in planes perpendicular to $c$ at heights of $z=0$, $c/4$, $c/2$ and $3c/4$. For Ni(II) ions in the $z=0$ and $z=c/2$ planes, the axial direction of the local octahedra is orientated along the $[\Bar{1}\,1\,0]$ direction. Owing to the $4_{1}$ screw symmetry, this axial direction is rotated by $90^{\circ}$ to align along $[1\,1\,0]$ for Ni(II) ions in the $z=c/4$ and $z=3c/4$ planes. This rotation ensures if a SIA anisotropy is present in these systems it will be rotated by $90^{\circ}$ between neighbouring spins connected by pym.

Pym ligands have been shown to be an effective mediator of exchange interactions of the order of $10$\,K~\cite{XCl2L_Jem_2024}. Secondary interactions are mediated through the halide-halide exchange bond, shown as dashed lines in Fig.~\ref{fig: NiX2_struct} (a) and (b). Such exchange paths depend strongly on the halide ions but are known to carry exchange of the order of $1$\,K~\cite{Blackmore-correlations}. This results in weakly interacting Ni---$X\cdot\cdot\cdot X$---Ni chains propagating along the $[\bar{1}\,1\,0]$ direction for ions in the $z=0$ and $z=c/2$ planes and along $[1\,1\,0]$ for ion in the $z=c/4$ and $z=3c/4$ planes. In the cases where there is AFM coupling through both the pym ligands and the halide-halide exchange bonds, the two interactions will be in competition, resulting in a reduced net effective interaction.

The spin Hamiltonian of Ni$X_{2}$(pym)$_{2}$, written in terms of the two spin sites $i$ and $j$ is,
\begin{dmath}
   \mathcal{H}= J\sum_{\left \langle i,j \right \rangle}\mathbf{S}_{i}\cdot\mathbf{S}_{j}
    +\sum_{(i,j)}g\mu_{\text{B}}\mu_{0}\mathbf{H}\cdot\mathbf{S}_{i,j}.
     +D \left[\sum_{i}\left (\hat{\mathbf{D}}_{i}\cdot\mathbf{S}_{i} \right )^{2}
     +\sum_{j} \left (\hat{\mathbf{D}}_{j}\cdot\mathbf{S}_{j} \right )^{2} \right ]
    \label{eq: Hamitonian}   
\end{dmath}
Here $\mathbf{S}_{i,j}$ is the spin of each ion, $\left \langle i,j \right \rangle$ denotes the sum over unique nearest neighbour exchange bonds with interaction strength $J$, $D$ is the size of the SIA, $\hat{\mathbf{D}_{i}} = [\bar{1} 1 0]/\sqrt{2}$ and $\hat{\mathbf{D}}_{j} = [1 1 0]/\sqrt{2}$ are the hard-axis unit vectors for sites $i$ and $j$, and $g\mu_{\text{B}}\mu_{0}\mathbf{H}\cdot\mathbf{S}_{i,j}$ is the Zeeman energy term assuming isotropic $g$-factor.

\subsection{Magnetometry}
\label{sec: mag}
\begin{figure}
    \centering
    \includegraphics[width= \linewidth]{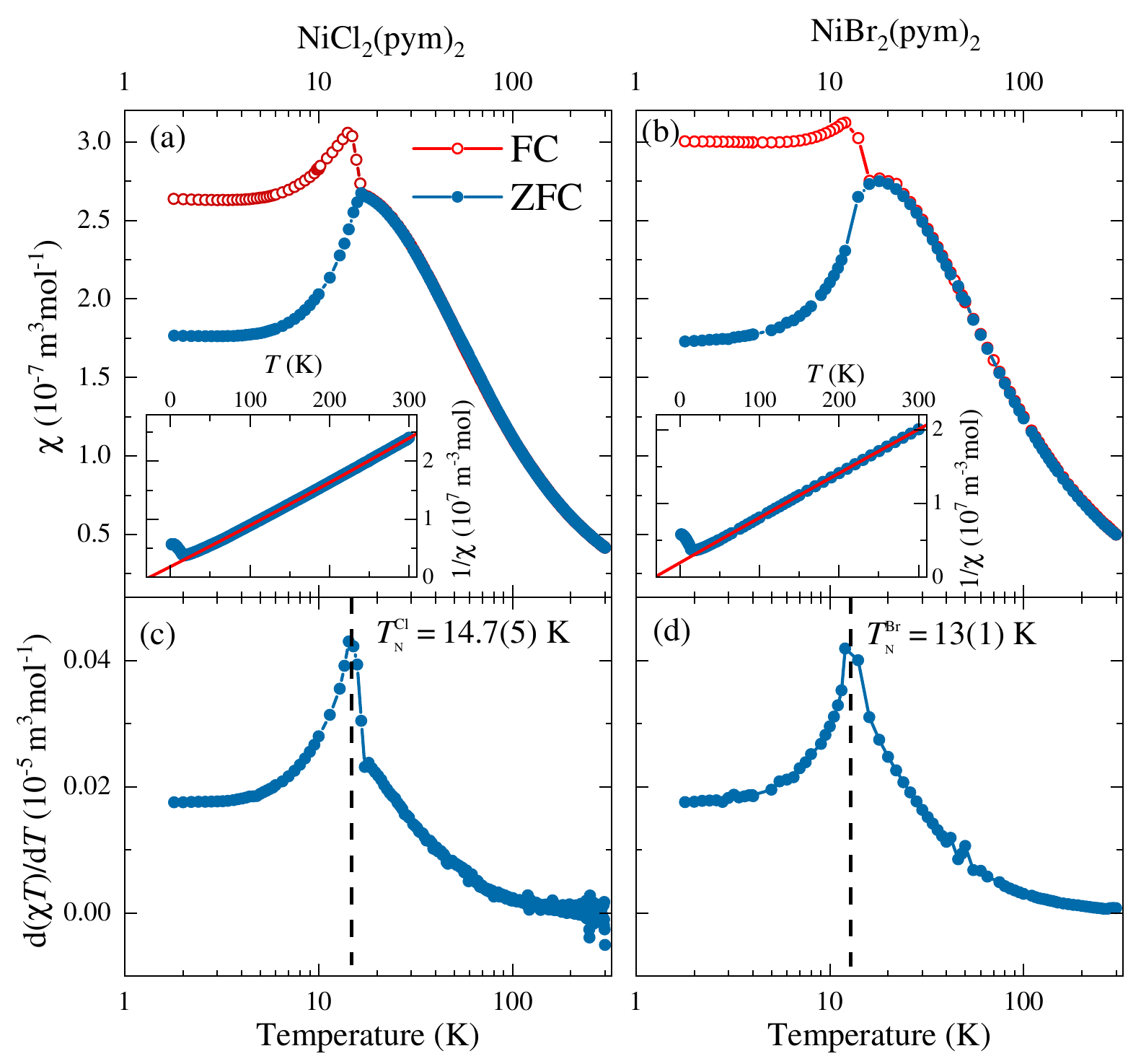}
    \caption{ Temperature dependent magnetic susceptibility $\chi(T)$ for powder samples of Ni$X_{2}$(pym)$_{2}$, where (a) $X$\,=\,Cl and (b) $X$\,=\,Br. For both a field of $\mu_{0}H = 0.1$\,T was applied, the field-cooled data are depicted as red circles and zero-field-cooled (ZFC) data are blue circles. The inset presents the Curie-Weiss fit to the ZFC $\chi^{-1}$. (c) and (d) Peak in the ZFC $d(\chi T)/dT$ marks the onset of magnetic long-range order at $T_{\text{N}} = 14.7(5)$\,K for $X$\,= Cl and $T_{\text{N}} = 13(1)$\,K for $X$\,= Br.}
    \label{fig: NiX2_chi}
\end{figure}

Powder $\chi(T)$ data for $X$\,=\,Cl are presented in Fig.~\ref{fig: NiX2_chi}(a). On cooling from $300$\,K, both the field-cooled (FC) and zero-field-cooled (ZFC) data exhibit a smooth rise to a sharp cusp and a bifurcation of the two curves at $T\approx15$\,K. Plotting the ZFC data as $d(\chi T)/dT$ in Fig.~\ref{fig: NiX2_chi}(d) reveals a lambda-like peak, close to the bifurcation, which is suggestive of a transition to long-range magnetic order below $T_{\rm{N}} = 14.7(5)$\,K~\cite{fisher_1962}. Further cooling results in plateaus in the FC and ZFC curves below $4.6$\,K. In the inset to Figure~\ref{fig: NiX2_chi}(a), a Curie-Weiss (CW) fit to the inverse susceptibility $\chi(T)^{-1}$, for $50$\,K$\leq T \leq 300$\,K, yields $g = 2.14(1)$, CW temperature $\theta_{\text{CW}} = -26.1(2)$\,K and a temperature independent constant $\chi_{0}=-2.62(5) \times 10^{-9}$\,m$^{3}$mol$^{-1}$. The negative $\theta_{\text{CW}}$ here indicates an AFM coupling of the Ni ions.

The $\chi(T)$ curves for $X$\,=\,Br in Fig.~\ref{fig: NiX2_chi}(b) displays a similar behaviour to $X$\,=\,Cl. However, the bifurcation and the lambda peak in ZFC  $d(\chi T)/dT$ [see Fig.~\ref{fig: NiX2_chi}(d)] are suppressed to a lower $T_{\text{N}} = 13(1)$\,K revealing a rounded local maximum prior to the bifurcation of the ZFC and FC curves. CW fits to the $\chi(T)^{-1}$ [see inset to Fig.~\ref{fig: NiX2_chi}(b)], results in  $g = 2.27(1)$, $\theta_{\text{CW}} = -31.7(2)$\,K and $\chi_{0}=1.09(9) \times 10^{-9}$\,m$^{3}$mol$^{-1}$.

\begin{figure}
    \centering
    \includegraphics[width= \linewidth]{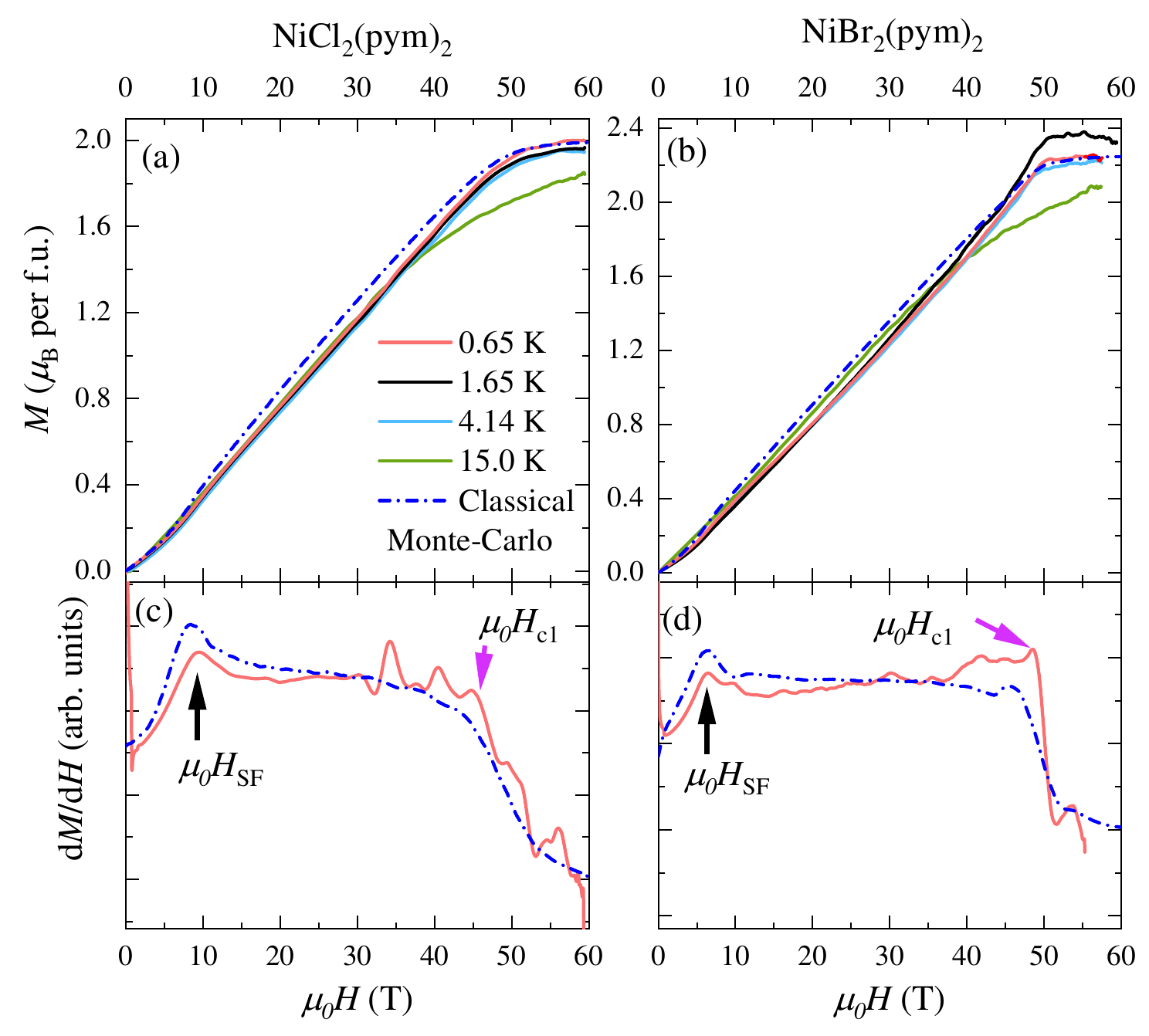}
    \caption{Powder-averaged magnetization, $M(H)$, data for Ni$X_{2}$(pym)$_{2}$, where (a) $X$\,=\,Cl and (c) $X$\,=\,Br at various temperatures. Differential susceptibility of the magnetization, d$M$/d$H$, below ordering temperature for (b) $X$\,=\,Cl and (d) $X$\,=\,Br. The black arrows mark the spin-flop field in both and the blue dash-dotted lines show the results of the classical Monte-Carlo simulation discussed in Section~\ref{sec: calcs} of the text.}
    \label{fig: NiX2_M}
\end{figure}

Figure~\ref{fig: NiX2_M}(a) and (b) present the pulsed-field $M(H)$ data for $X$\,=\,Cl and Br respectively. In both materials, at low fields and temperatures below $T_{\text{N}}$, there is an upturn in the magnetization observed as a peak in the differential susceptibility, d$M$/d$H$, shown in Fig.~\ref{fig: NiX2_M}(c) and (d). These features, reminiscent of the spin-flop transition found in easy-axis systems, are observed at $\mu_{0}H_{\text{SF}} = 9.5(8)$\,T and $6.6(8)$\,T for $X$\,=\,Cl and Br respectively. In the $X$\,=\,Cl sample, further increase in field results in a near linear rise in $M(H)$, up to the first saturation field $H_{\text{c}1} = 46(3)$\,T. $H_{\text{c}1}$ is defined as the point where $M(H)$ deviates from a linear field-dependence (point where d$M$/d$H$ starts to decrease). The $0.65$\,K $M(H)$ data for $X$\,=\,Br display a slightly concave approach to the first saturation field, resulting in a peak in d$M$/d$H$ at $H_{\text{c}1} = 48(1)$\,T. As we will show in Section~\ref{sec: calcs}, $H_{\text{c}1}$ corresponds to magnetic saturation for grains in which the field aligns along the $[0\,0\,1]$ axis. For $H>H_{\text{c}1}$, the projected value of the saturation magnetization suggests a low-temperature $g$ factor of $g = 2.06$ for $X$\,=\,Cl and $g = 2.25$ for $X$\,=\,Br. The $g$ for $X$\,=\,Cl is slightly lower than predicted for the CW fit at high temperatures while the values for $X$\,=\,Br are in good agreement.

For a spin-canted magnetic structure, similar to FeCl$_{2}$(pym)$_{2}$, a remanent $M(H)$ at zero-field is expected. However in both $X$\,=\,Cl and Br, very small remanent $M(0)$ of $0.001$\,$\mu_{\text{B}}$ and $0.00016$\,$\mu_{\text{B}}$\,per Ni(II) ion respectively, are observed~\cite{supplementary}. These small $M(0)$ are readily explained by the presence of a small amount of Ni$X_{2}$(pym) impurity phases (as seen in the crystallographic data for the $X$\,=\,Cl sample in the supplementary material~\cite{supplementary}) which are known to exhibit a $M(0)=0.127$\,$\mu_{\text{B}}$\,per\,Ni(II)~\cite{XCl2L_Jem_2024}.

\subsection{Muon-spin rotation}
\label{sec: muon}
\begin{figure}
    \centering
    \includegraphics[width= \linewidth]{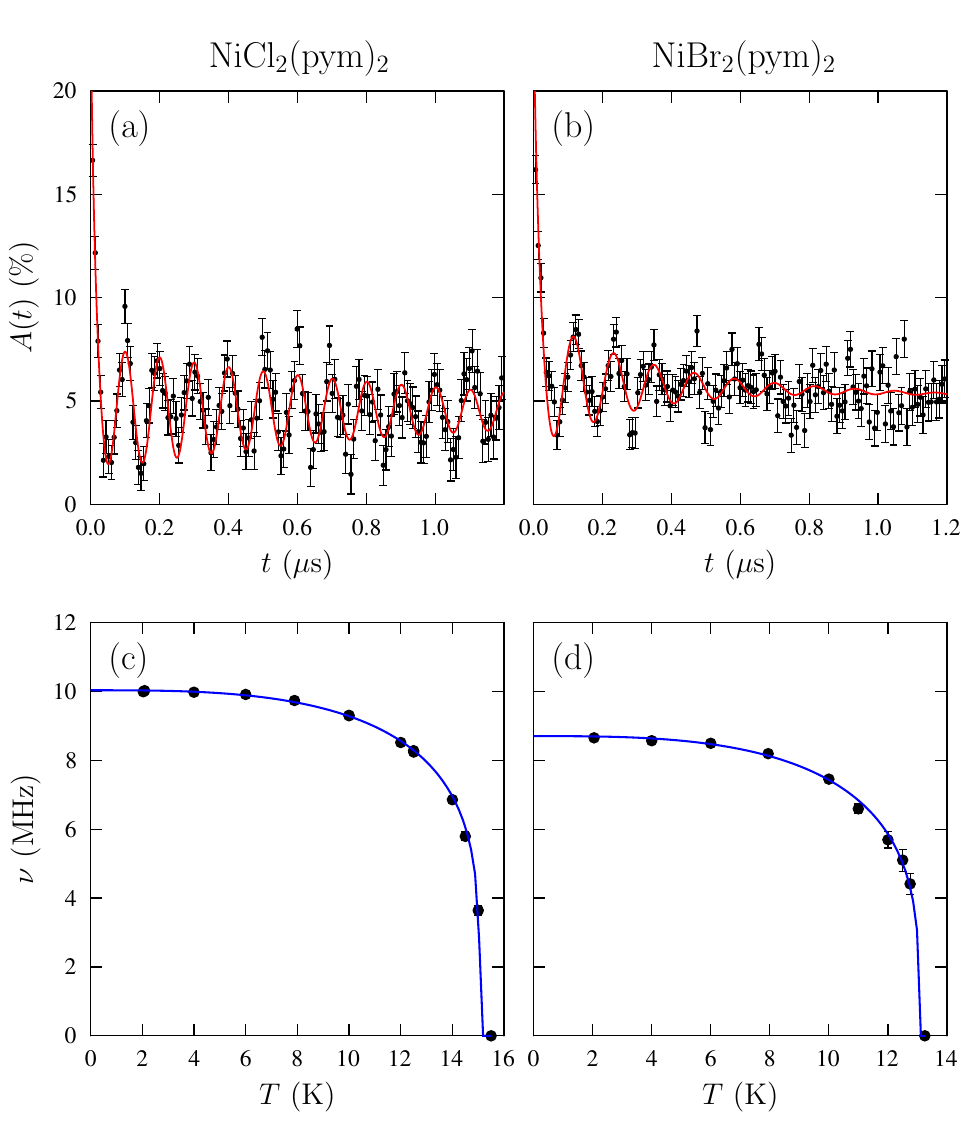}
    \caption{ The zero-field $\mu^{+}$SR spectra data (black circles) collected at $T = 4$\,K for Ni$X_{2}$(pym)$_{2}$, where (a) $X$\,=\,Cl and (b) $X$\,=\,Br. The temperature dependence of the oscillation frequency $\nu$ for (c) $X$\,=\,Cl and (d) $X$\,=\,Br. The solid lines represent fits to the data.} 
    \label{fig: Muon}
\end{figure}

Zero-field $\mu^{+}$SR measurements were carried out at the Swiss Muon Source, Paul Scherrer Institut. Data were collected at temperatures $2 \leq T \leq 16$\,K and example spectra at $4$\,K for $X$\,=\,Cl and $X$\,=\,Br are shown in Fig.~\ref{fig: Muon}  (a) and (b) respectively. For both materials, the time-dependent asymmetry function exhibits clear oscillations, indicative of long-range magnetic order throughout the bulk of the material. The quasistatic local field, arising in the presence of long-range magnetic order, causes a coherent precession of the spin of those muons for which a component of their spin polarization lies perpendicular to this local field (expected to be 2/3 of the total spin polarization for a polycrystalline sample). The frequency of the oscillation is given by $\nu_{i} = \gamma_{\mu}|B_{i}|/2\pi$, where $\gamma_{\mu} = 2\pi \times 135.5$\, MHz T$^{-1}$ is the muon gyromagnetic ratio and $B_{i}$ is the average magnitude of the local magnetic field at the $i$th muon site. As such $\nu_{i}$ is an effective order parameter and indicative of the total ordered moment within each sample. 

To describe the temperature evolution of the asymmetry function and hence $\nu_{i}(T)$, fits were produced to a function of the form
\begin{equation}
    A(t) = A_{1}e^{-\lambda_{1}t}\cos(2\pi\nu t) + A_{2}e^{-\lambda_{2}t} + A_{\text{b}},
\label{eq: A(t)}
\end{equation}
where $\lambda_{i}$ are relaxation rates and $A_{\text{b}}$ is a background contribution from muons stopping outside the sample. The first term in Eq.~\ref{eq: A(t)} accounts for the oscillatory signal and the second term accounts for the muon polarization initially parallel to the local magnetic field. The presence of a single frequency for both materials implies that a single muon site gives rise to the oscillatory components of the spectra. The resulting temperature-dependent $\nu(T)$, shown in Fig.~\ref{fig: Muon}(c) and (d), is described using the phenomenological function
\begin{equation}
    \nu(T)=\left [1 - \left ( \frac{T}{T_{N}} \right )^{\alpha}  \right ]^{\beta}.
\label{eq: freq(T)}
\end{equation}

For $X$\,=\,Cl, we obtain $\nu(0)=10.0(1)$\,MHz  (corresponding to a local field of $74$\,mT), $T_{\text{N}} = 15.05(1)$ and $\beta = 0.22(1)$ for a fixed $\alpha = 3$. The ordering temperature extracted here is in excellent agreement with the analysis of the $\chi(T)$ data. The rapid drop in $\nu(T)$ with increasing temperature along with the fitted value of $\beta$ implies the existence of fluctuations, close to the phase transition, that are intermediate between a two- and three-dimensional character. For $X$\,=\,Br, the fit yields $\nu(0)=8.70(3)$\,MHz  (corresponding to a local field of $64$\,mT), $T_{\text{N}} = 13.1(1)$ and $\beta = 0.27(1)$. The ordering temperature is again in excellent agreement with the $\chi(T)$ data. Compared to $X$\,=\,Cl, an increased $\beta$ in $X$\,=\,Br implies the fluctuations here are more three-dimensional. This is in keeping with the expectation that there are non-zero exchange interactions through the halides in the $X$\,=\,Br system, while in $X$\,=\,Cl this interaction is expected to be much smaller~\cite{Blackmore-correlations}.

\subsection{Elastic neutron scattering}
\label{sec: neutron}
\begin{figure}[h]
    \centering
    \includegraphics[width= \linewidth]{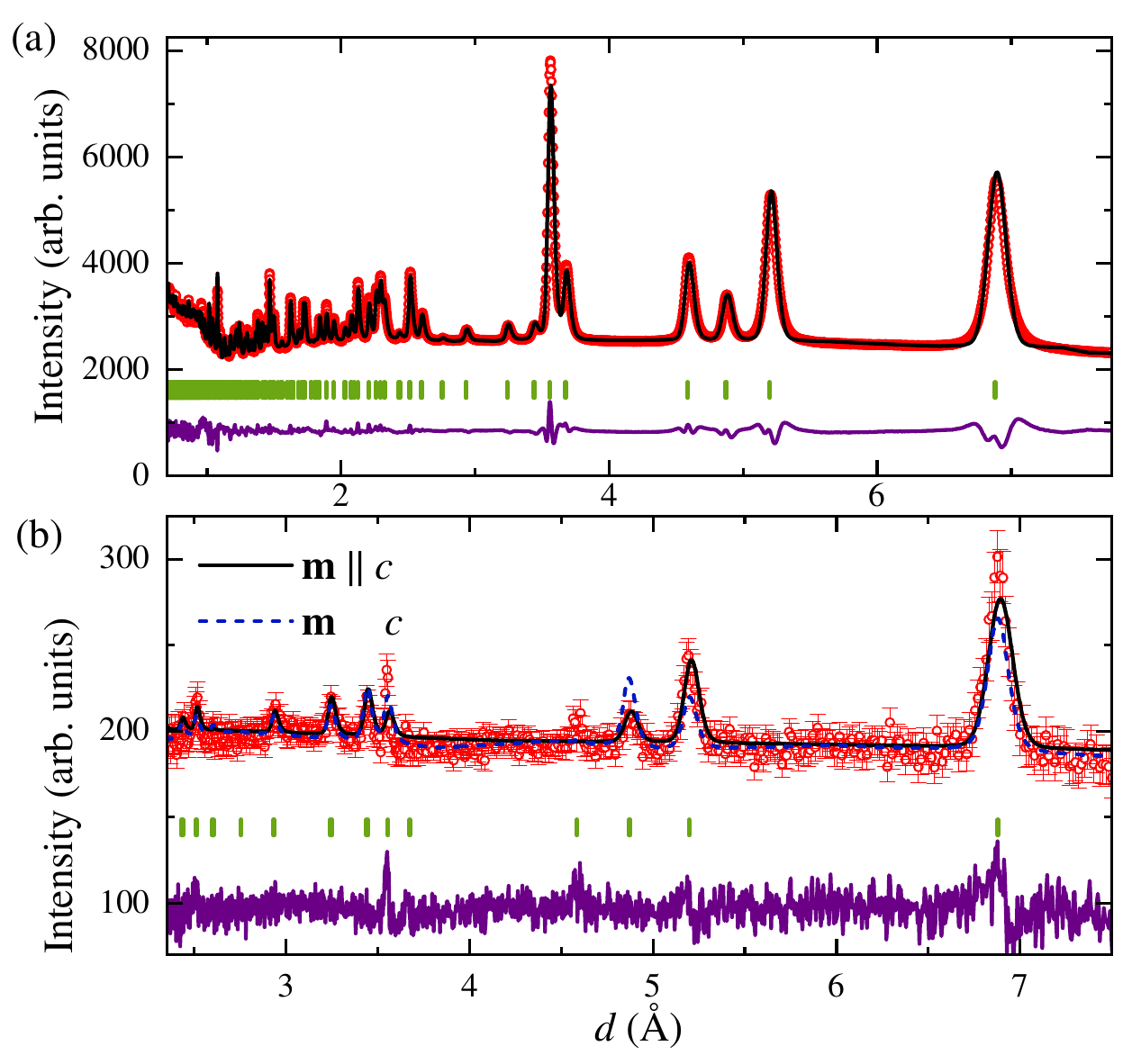}
    \caption{(a) Rietveld refinement of the $I4_{1}22$ the nuclear structure of NiCl$_{2}$(pym)$_{2}$ to powder neutron diffraction data collected at $T=20$\,K. The data points are shown as red circles, fitting curve in black, Bragg positions are marked by green ticks and the difference between the data and the calculated pattern is shown in purple. (b) Magnetic diffraction pattern obtained by subtracting data collected at $20$\,K from the $1.4$\,K data. The fitted pattern (black) has magnetic moments parallel to the $c$-axis compared to the pattern with moments perpendicular to $c$ (blue dashed line).}
    \label{fig: NiCl2_elastic}
\end{figure}

Elastic neutron diffraction measurements on powder samples of NiCl$_{2}$(pym)$_{2}$ were performed on the WISH instrument at the ISIS, UK Neutron and Muon Source~\cite{Wish}, at temperatures above ($T = 20$\,K) and below ($T = 1.4$\,K) the magnetic ordering transition. Rietveld refinement of the nuclear structure was performed on the data collected at $T = 20$\,K, across four separate detector banks, using the FULLPROF software package~\cite{Fullprof}. Figure~\ref{fig: NiCl2_elastic}(a) present the structural fit with $R_{\text{Bragg}} = 3.361 \%$ to data from one such detector bank at average $2\theta=58.330^{\circ}$. The resulting lattice parameters,  $a = 7.3425(2)\,\si{\angstrom}$ and $c = 19.4841(9)\,\si{\angstrom}$ and nuclear structure are in good agreement with the values determined from the electron diffraction and PXRD measurements at $120$\,K and accounted for all observed peaks at $20$\,K.

The difference between the data collected at $T=5$\,K and $T=20$\,K reveals magnetic Bragg peaks residing on top of the nuclear peak positions, indicating magnetic order with propagation vector $\textbf{k}=(0,0,0)$. Symmetry analysis in ISODISTORT~\cite{Isodistort,Isodistort_2006} reveals three candidate symmetries (irreps $m\Gamma_{2}$, $m\Gamma_{3}$ and $m\Gamma_{5}$) for the magnetic structure. Among these, $m\Gamma_{3}$ corresponds to ferromagnetic (FM) order and is ruled out from the $\chi(t)$ measurements. The magnetic moments in the magnetic structure described by the $m\Gamma_{2}$ irrep are constrained to point parallel to the $c$-axis with nearest neighbours antiferromagnetically coupled through pym. Consequently, the ordered moment size is the only refinable parameter in this irrep. The structure associated with $m\Gamma_{5}$ irrep decomposes into a linear combination of perpendicular FM and AFM modes within the $ab$-plane and hence encompasses the canted structure observed in the FeCl$_{2}$(pym)$_{2}$ system~\cite{FeCl2}.
\begin{figure}
    \centering
    \includegraphics[width= \linewidth]{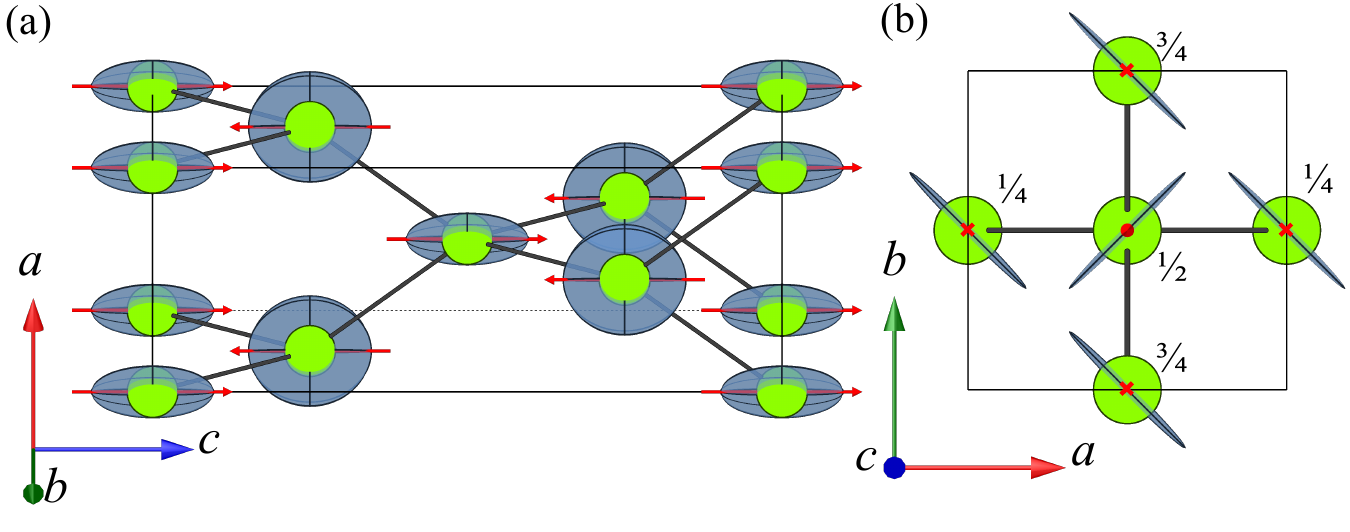}
    \caption{(a) Zero-field magnetic structure of NiCl$_{2}$(pym)$_{2}$ determined through powder neutron diffraction. The red arrows indicate the magnetic moment direction of Ni(II) ions (green spheres). The blue circles show the local single-ion easy plane of Ni(II) ions and the black lines show the nearest neighbours' exchange interactions. (b) View down the $c$-axis illustrates 90$^{\circ}$ rotation of the easy-plane between nearest neighbours. Ni(II) ions at $[x,y,0]$ and $[x,y,1]$ have been omitted for clarity.}
    \label{fig: NiCl2_mag_strut}
\end{figure}

The diffraction pattern resulting from the refinement of both AFM models is shown in Fig.~\ref{fig: NiCl2_elastic}(b). The magnetic structure associated with irrep $m\Gamma_{2}$ is found to best fit the data with $R_{\text{mag}} =  4.16 \%$ compared to $R_{\text{mag}} =  34.89 \%$ of irrep $m\Gamma_{5}$. The resulting ordered moment size is $1.99(3)$\,$\mu_{\text{B}}$ per Ni(II) ion, which is in excellent agreement with the low-temperature magnetic saturation value. The resulting magnetic structure is depicted in Fig.~\ref{fig: NiCl2_mag_strut} and is consistent with an easy-plane SIA. While the local easy-plane, defined by the local Ni---pym equatorial plane and shown as blue circles in Fig.~\ref{fig: NiCl2_mag_strut}, rotates by $90^{\circ}$ between each neighbouring Ni(II) ions, the intersection of these planes defines a pseudo-easy-axis along the $c$ direction. Moments aligning along this axis satisfy both the exchange interaction through pym and the SIA. Additionally, the $\textbf{k}=(0,0,0)$ vector and the magnetic structure is only possible when the exchange through the pym is dominant and AFM.

Our bulk magnetic characterisation suggests X =Cl and X =Br are described by the same magnetic Hamiltonians but with differing values of $J$ and $D$. Since the magnetic ground state determined for $X$\,=\,Cl is independent of the values of AFM $J$ and easy-plane $D$, we expect NiBr$_{2}$(pym)$_{2}$ will adopt the same magnetic structure.

\subsection{Calculations and Simulations}
\label{sec: calcs}
\subsubsection{Critcal Fields}
\begin{figure}
    \centering
    \includegraphics[width= \linewidth]{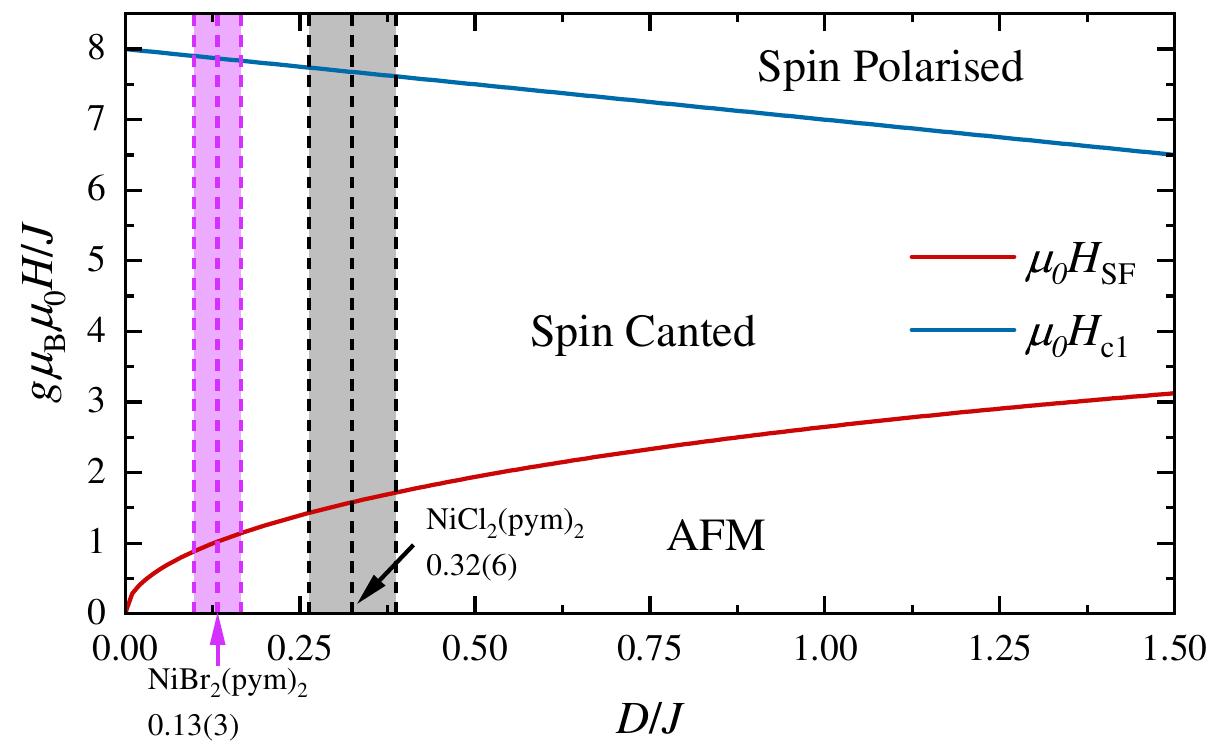}
    \caption{Phase diagram for Ni$X_{2}$(pym)$_{2}$, for $\mathbf{H}||c$ and calculated using a classical mean-field model described in the text. The $x$-axis is the ratio of easy-plane single-ion anisotropy and the exchange energies. The $y$-axis is the ratio of the Zeeman and the exchange energy. The red line represents the calculated spin-flop transition taking the systems from an antiferromagnetic state to a spin-canted state. The blue line is the boundary between the spin canted to the spin-polarised state. The black and purple dashed lines mark the $D/J$ ratio of the $X$\,=\,Cl and Br materials respectively and the shaded regions indicate the experimental uncertainty in these values.}
    \label{fig: NiCl2_Phase}
\end{figure}

In a similar manner to ordinary easy-axis systems, applying the field parallel to the pseudo-easy-axis will cause a spin-flop and magnetic saturation transitions which are seen in the $M(H)$ data in Fig~\ref{fig: NiX2_M}. Mean-field calculations, under the assumption that $D\ll nJ$, presented in the supplementary material~\cite{supplementary}, predict that these transitions occur at,
\begin{dmath}
   \mu_{0}H_{\text{c}1} = \frac{2nJ -D}{g\mu_{B}},
   \label{eq: H_z_sat}   
\end{dmath}
\begin{dmath}
    \mu_{0}H_{\text{SF}} = \frac{\sqrt{2nJ D-D^{2}}}{g\mu_{B}}.
    \label{eq: Spin-flop}   
\end{dmath}
Here $n=4$ is the number of nearest neighbours. Taking the isotropic low-temperature $g$-factors and the critical field from the $M(H)$ data, we determine that $J = 8.1(5)$\,K, $D = 2.6(5)$\,K for $X$\,=\,Cl and $J = 9.2(2)$\,K, $D = 1.2(3)$\,K for $X$\,=\,Br. The phase diagram of $D/J$ ratio against the ratio of Zeeman energy and exchange energy $g\mu_{\text{B}}\mu_{0}H/J$ is shown in Fig.~\ref{fig: NiCl2_Phase}. The $D/J$ ratios for both materials have been marked in the phase diagram and are summarised in Table~\ref{tab: J_D}.
\begin{table}
\caption{\label{tab: J_D} The AFM exchange energy, $J$, and the easy-plane single-ion anisotropy parameter, $D$, derived using mean-field calculations of the observed critical fields in the magnetization, $M(H)$, data. The $g$-factors are the low-temperature values determined experimentally through the $M(H)$ saturation value.}
\begin{ruledtabular}
\begin{tabular}{cccc}
\textrm{}&
\textrm{$J$\,(K)}&
\textrm{$D$\,(K)}&
\textrm{$g$}\\
\colrule
NiCl$_{2}$(pym)$_{2}$ & $8.1(5)$  & $2.6(5)$ & $2.06$ \\
NiBr$_{2}$(pym)$_{2}$ & $9.2(2)$  & $1.2(3)$ & $2.25$ \\
\end{tabular}
\end{ruledtabular}
\end{table}


\subsubsection{Monte-Carlo simulations}
\begin{figure}
    \centering
    \includegraphics[width= \linewidth]{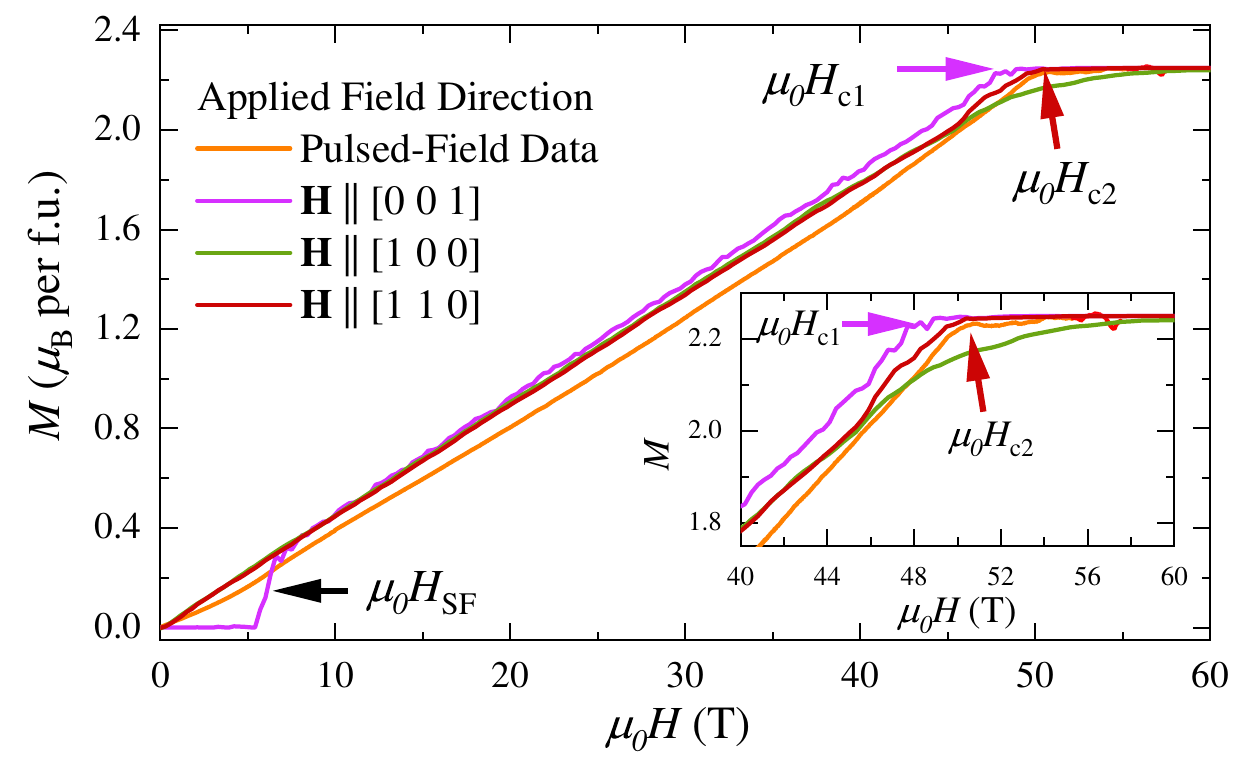}
    \caption{Monte-Carlo simulation of the magnetization $M(H)$ for NiBr$_{2}$(pym)$_{2}$ with the field along various crystallographic directions along with the measured pulsed field $M(H)$ data. The inset is zoomed into the $40\geq\mu_{0}H\geq60$\,T range. The black arrow highlights the spin-flop field and the purple arrow points toward the saturation field for $\mathbf{H} || $[0\,0\,1] (field parallel to pseudo-easy-axis). The red arrow indicates the saturation field for $\mathbf{H} || $[1\,1\,0](perpendicular to the pseudo-easy-axis but in the easy-plane of a sublattice). There is no magnetic saturation in the $M(H)$ for fields that are not aligned along the easy-plane of either sublattice in the systems (green line)}
    \label{fig: NiBr2_MC}
\end{figure}

To confirm the values of $J$ and $D$, determined using Eq.~\ref{eq: H_z_sat} and~\ref{eq: Spin-flop} and reported in Table.~\ref{tab: J_D}, classical MC simulations of the powder-averaged $M(H)$ were carried out. $M(H)$ curves were calculated for 100 different field directions evenly spaced on a unit sphere and averaged to calculate the powder averaged $M(H)$. The resulting simulated $M(H)$ curves for $X$\,=\,Cl and Br are shown in Fig.~\ref{fig: NiX2_M} (a) and (b) respectively and the differential susceptibilities are shown in Fig.~\ref{fig: NiX2_M} (c) and (d).

In both $X$\,=\,Cl and Br, there is a good agreement between the experimental and the simulated $M(H)$ curves. The simulations show a small overestimation of the $M(H)$ due to the slight concavity in the measured $M(H)$ curves. In [Ni(HF$_{2}$)(pyz)$_{2}$]SbF$_{6}$, a similar discrepancy was attributed to the development of quantum fluctuations resulting from the quasi-one-dimensional nature of the magnetic exchange~\cite{Brambleby_2017}. Here the discrepancy may also arise from quantum fluctuations in these $S=1$ systems. 

For $X$\,=\,Br, there is a good agreement between simulated and observed $\mu_{0}H_{\text{SF}}$ and $\mu_{0}H_{\text{c}1}$. For $X$\,=\,Cl, the larger $D/nJ$ results in a slight underestimation of $D$ when calculated using Eq.~\ref{eq: Spin-flop}, which is derived under the approximation that $D\ll nJ$. Consequently, MC simulations using the calculated $J$ and $D$ values (as reported in Table.~\ref{tab: J_D}) results in a slightly smaller simulated $\mu_{0}H_{\text{SF}}$ than the observed value for $X$\,=\,Cl. However, as seen in Fig.~\ref{fig: NiX2_M} (a) and (c), the deviation in $\mu_{0}H_{\text{SF}}$ is small and the derived value of $D$ is expected to be correct within the experimental error.

 It is often the case that for anisotropic magnets, a symmetry-breaking phase transition occurs at saturation when a sufficiently large magnetic field is applied along the principle anisotropy axes. In our systems, the rotational symmetry around the SIA axis is broken by the alternation of the Ni(II) octahedra. As a result, field directions perpendicular to the pseudo-easy-axis exist for which a phase transition is absent at finite fields. Here, the SIA can always save energy by canting spins away from the applied field and a spin-polarized phase is only reached in the limit of infinite field. This is illustrated by the MC simulation of $M(H)$ for $\mathbf{H} || [1\,0\,0]$ presented in Fig.~\ref{fig: NiBr2_MC}, in which $M(H)$ asymptotically approaches the saturation value. When the applied field is parallel to the pseudo-easy-axis, the field-polarised phase minimises the SIA energy for spin in both $i$ and $j$ sites resulting in the first saturation phase transition at $\mu_{0}H_{\text{c}1}$. A second saturation field is expected to occur for $\mathbf{H} || [1\,1\,0]$ or $[\bar{1}\,1\,0]$, corresponding to direction in which the field aligns parallel to the hard-axis of one site and also within the easy-plane of the other site~\footnote{The hard-axis of one site aligns within the easy-plane of the second site because of the exact $90^{\circ}$ rotation of the Ni(II) between neighbouring sites. For rotation by an angle which is not a multiple of $90^{\circ}$, saturation phase transitions will not occur for any directions perpendicular to the pseudo-easy-axis.}. In a powder-averaged $M(H)$ experiment, the second saturation field is hard to define in the data.

\section{Conclusions}
\label{sec: Conclusion}

In summary, we have thoroughly characterised the magnetic properties of the $S=1$ diamond-lattice systems, Ni$X_{2}$(pym)$_{2}$ ($X$\,=\,Cl or Br). The octahedral orientation and the SIA axis of neighbouring Ni(II) ions connected by AFM exchange-mediating pym ligands alternate by a $90^{\circ}$ rotation about the $[0\,0\,1]$ direction. While an alternating easy-axis induces a spin-canted magnetic order, as seen in FeCl$_{2}$(pym)$_{2}$~\cite{FeCl2}, NiCl$_{2}$(pym) and NiCl$_{2}$(btd)~\cite{XCl2L_Jem_2024}, an alternating easy-plane anisotropy is expected to result in collinear order along the unique direction created by the intersection of the two easy-planes. Here we denote this direction as the pseudo-easy-axis. Recent studies on FeCl$_{2}$(pym) and FeCl$_{2}$(btd) found that while hosting an alternating easy-plane anisotropy, significant spin-canting arose due to rhombic anisotropy, DM or higher order interaction~\cite{XCl2_pressure}. In contrast, our neutron diffraction measurements reveal that the magnetic moments in NiCl$_{2}$(pym)$_{2}$ align collinearly along $[0\,0\,1]$, the pseudo-easy-axis direction.

Similar $\chi(T)$ and a low-field feature in $M(H)$, which is ascribed to a spin-flop transition, is observed in both $X$\,=\,Cl and Br compounds, implying that $X$\,=\,Br also hosts a pseudo-easy-axis and adopts the same collinear order. Mean-field calculations of the spin-flop and the magnetization saturation field were used to determine that $J = 8.1(5)$\,K, $D = 2.6(5)$\,K for $X$\,=\,Cl and $J = 9.2(2)$\,K, $D = 1.2(3)$\,K for $X$\,=\,Br. The values of $J$ calculated here are comparable to $J=7$\,K reported in Ref.~\cite{XCl2_pressure} and MC simulations of the powder-average $M(H)$ simulations were performed to verify these Hamiltonian parameters. Moreover, MC simulations of single-crystal $M(H)$ curves indicate that when the applied field is parallel to the pseudo-easy-axis, characteristic spin-flop and a magnetic saturation phase-transition, akin to conventional easy-axis systems, occur. However, due to the broken rotational symmetry around the pseudo-easy-axis, there are field directions perpendicular to the pseudo-easy-axis where the $M(H)$ exhibits an asymptotic approach to magnetic saturation rather than a saturation phase transition at finite fields.

The reduction in $D$ and an increase in secondary exchange through the halide-halide bond in $X$\,=\,Br along is in keeping with the magneto-structural correlations established for halides of different radii in Ni$X_{2}$(3,5-lut)$_{4}$~\cite{Blackmore-correlations}. Similar to NiI$_{2}$(3,5-lut)$_{4}$, we expect NiI$_{2}$(pym)$_{2}$ to exhibit a small easy-axis $D$ accompanied by a larger halide-halide exchange interaction. This would lead to the observation of a SIA-induced spin-canted magnetic order, similar to that observed in isostructural FeCl$_{2}$(pym)$_{2}$~\cite{FeCl2}.

\begin{acknowledgments}
We are indebted to the late Jamie Manson for instigating this work, for his role in designing and growing the samples and for many other invaluable contributions. We thank T. Orton and P. Ruddy for technical assistance. SV thanks the EPSRC (UK) for support. This project has received funding from the European Research Council (ERC) under the European Union’s Horizon 2020 research and innovation programme (grant agreement No. 681260). A portion of this work was performed at the National High Magnetic Field Laboratory, which is supported by National Science Foundation Cooperative Agreements Nos. DMR-1644779 and DMR-2128556, the US Department of Energy (DoE) and the State of Florida. JS acknowledges support from the DoE BES FWP “Science of 100 T". AHM thanks the STFC and EPSRC for supporting his studentship. Part of this work was carried out at the Swiss Muon Source, Paul Scherrer Institut and we are grateful for the provision of beamtime. The authors also thank the EPSRC for funding (EP/X014606/1, A National Electron Diffraction Facility for Nanomaterial Structural Studies), and the University of Warwick X-ray Research Technology Platform for the provision of further analysis facilities. Data presented in this paper will be made available at XXX
\end{acknowledgments}
\bibliography{main.bib}

\begin{thebibliography}{40}%
\makeatletter
\providecommand \@ifxundefined [1]{%
 \@ifx{#1\undefined}
}%
\providecommand \@ifnum [1]{%
 \ifnum #1\expandafter \@firstoftwo
 \else \expandafter \@secondoftwo
 \fi
}%
\providecommand \@ifx [1]{%
 \ifx #1\expandafter \@firstoftwo
 \else \expandafter \@secondoftwo
 \fi
}%
\providecommand \natexlab [1]{#1}%
\providecommand \enquote  [1]{``#1''}%
\providecommand \bibnamefont  [1]{#1}%
\providecommand \bibfnamefont [1]{#1}%
\providecommand \citenamefont [1]{#1}%
\providecommand \href@noop [0]{\@secondoftwo}%
\providecommand \href [0]{\begingroup \@sanitize@url \@href}%
\providecommand \@href[1]{\@@startlink{#1}\@@href}%
\providecommand \@@href[1]{\endgroup#1\@@endlink}%
\providecommand \@sanitize@url [0]{\catcode `\\12\catcode `\$12\catcode `\&12\catcode `\#12\catcode `\^12\catcode `\_12\catcode `\%12\relax}%
\providecommand \@@startlink[1]{}%
\providecommand \@@endlink[0]{}%
\providecommand \url  [0]{\begingroup\@sanitize@url \@url }%
\providecommand \@url [1]{\endgroup\@href {#1}{\urlprefix }}%
\providecommand \urlprefix  [0]{URL }%
\providecommand \Eprint [0]{\href }%
\providecommand \doibase [0]{http://dx.doi.org/}%
\providecommand \selectlanguage [0]{\@gobble}%
\providecommand \bibinfo  [0]{\@secondoftwo}%
\providecommand \bibfield  [0]{\@secondoftwo}%
\providecommand \translation [1]{[#1]}%
\providecommand \BibitemOpen [0]{}%
\providecommand \bibitemStop [0]{}%
\providecommand \bibitemNoStop [0]{.\EOS\space}%
\providecommand \EOS [0]{\spacefactor3000\relax}%
\providecommand \BibitemShut  [1]{\csname bibitem#1\endcsname}%
\let\auto@bib@innerbib\@empty
\bibitem [{\citenamefont {Dender}\ \emph {et~al.}(1997)\citenamefont {Dender}, \citenamefont {Hammar}, \citenamefont {Reich}, \citenamefont {Broholm},\ and\ \citenamefont {Aeppli}}]{Dender_1997}%
  \BibitemOpen
  \bibfield  {author} {\bibinfo {author} {\bibfnamefont {D.~C.}\ \bibnamefont {Dender}}, \bibinfo {author} {\bibfnamefont {P.~R.}\ \bibnamefont {Hammar}}, \bibinfo {author} {\bibfnamefont {D.~H.}\ \bibnamefont {Reich}}, \bibinfo {author} {\bibfnamefont {C.}~\bibnamefont {Broholm}}, \ and\ \bibinfo {author} {\bibfnamefont {G.}~\bibnamefont {Aeppli}},\ }\href {\doibase 10.1103/PhysRevLett.79.1750} {\bibfield  {journal} {\bibinfo  {journal} {Phys. Rev. Lett.}\ }\textbf {\bibinfo {volume} {79}},\ \bibinfo {pages} {1750} (\bibinfo {year} {1997})}\BibitemShut {NoStop}%
\bibitem [{\citenamefont {Hong}\ \emph {et~al.}(2006)\citenamefont {Hong}, \citenamefont {Kenzelmann}, \citenamefont {Turnbull}, \citenamefont {Landee}, \citenamefont {Lewis}, \citenamefont {Schmidt}, \citenamefont {Uhrig}, \citenamefont {Qiu}, \citenamefont {Broholm},\ and\ \citenamefont {Reich}}]{Hong_2006}%
  \BibitemOpen
  \bibfield  {author} {\bibinfo {author} {\bibfnamefont {T.}~\bibnamefont {Hong}}, \bibinfo {author} {\bibfnamefont {M.}~\bibnamefont {Kenzelmann}}, \bibinfo {author} {\bibfnamefont {M.~M.}\ \bibnamefont {Turnbull}}, \bibinfo {author} {\bibfnamefont {C.~P.}\ \bibnamefont {Landee}}, \bibinfo {author} {\bibfnamefont {B.~D.}\ \bibnamefont {Lewis}}, \bibinfo {author} {\bibfnamefont {K.~P.}\ \bibnamefont {Schmidt}}, \bibinfo {author} {\bibfnamefont {G.~S.}\ \bibnamefont {Uhrig}}, \bibinfo {author} {\bibfnamefont {Y.}~\bibnamefont {Qiu}}, \bibinfo {author} {\bibfnamefont {C.}~\bibnamefont {Broholm}}, \ and\ \bibinfo {author} {\bibfnamefont {D.}~\bibnamefont {Reich}},\ }\href {\doibase 10.1103/PhysRevB.74.094434} {\bibfield  {journal} {\bibinfo  {journal} {Phys. Rev. B}\ }\textbf {\bibinfo {volume} {74}},\ \bibinfo {pages} {094434} (\bibinfo {year} {2006})}\BibitemShut {NoStop}%
\bibitem [{\citenamefont {Zapf}\ \emph {et~al.}(2006)\citenamefont {Zapf}, \citenamefont {Zocco}, \citenamefont {Hansen}, \citenamefont {Jaime}, \citenamefont {Harrison}, \citenamefont {Batista}, \citenamefont {Kenzelmann}, \citenamefont {Niedermayer}, \citenamefont {Lacerda},\ and\ \citenamefont {Paduan-Filho}}]{Zapf_2006}%
  \BibitemOpen
  \bibfield  {author} {\bibinfo {author} {\bibfnamefont {V.~S.}\ \bibnamefont {Zapf}}, \bibinfo {author} {\bibfnamefont {D.}~\bibnamefont {Zocco}}, \bibinfo {author} {\bibfnamefont {B.~R.}\ \bibnamefont {Hansen}}, \bibinfo {author} {\bibfnamefont {M.}~\bibnamefont {Jaime}}, \bibinfo {author} {\bibfnamefont {N.}~\bibnamefont {Harrison}}, \bibinfo {author} {\bibfnamefont {C.~D.}\ \bibnamefont {Batista}}, \bibinfo {author} {\bibfnamefont {M.}~\bibnamefont {Kenzelmann}}, \bibinfo {author} {\bibfnamefont {C.}~\bibnamefont {Niedermayer}}, \bibinfo {author} {\bibfnamefont {A.}~\bibnamefont {Lacerda}}, \ and\ \bibinfo {author} {\bibfnamefont {A.}~\bibnamefont {Paduan-Filho}},\ }\href {\doibase 10.1103/PhysRevLett.96.077204} {\bibfield  {journal} {\bibinfo  {journal} {Phys. Rev. Lett.}\ }\textbf {\bibinfo {volume} {96}},\ \bibinfo {pages} {077204} (\bibinfo {year} {2006})}\BibitemShut {NoStop}%
\bibitem [{\citenamefont {Brambleby}\ \emph {et~al.}(2017)\citenamefont {Brambleby}, \citenamefont {Goddard}, \citenamefont {Singleton}, \citenamefont {Jaime}, \citenamefont {Lancaster}, \citenamefont {Huang}, \citenamefont {Wosnitza}, \citenamefont {Topping}, \citenamefont {Carreiro}, \citenamefont {Tran}, \citenamefont {Manson},\ and\ \citenamefont {Manson}}]{Brambleby_2017}%
  \BibitemOpen
  \bibfield  {author} {\bibinfo {author} {\bibfnamefont {J.}~\bibnamefont {Brambleby}}, \bibinfo {author} {\bibfnamefont {P.~A.}\ \bibnamefont {Goddard}}, \bibinfo {author} {\bibfnamefont {J.}~\bibnamefont {Singleton}}, \bibinfo {author} {\bibfnamefont {M.}~\bibnamefont {Jaime}}, \bibinfo {author} {\bibfnamefont {T.}~\bibnamefont {Lancaster}}, \bibinfo {author} {\bibfnamefont {L.}~\bibnamefont {Huang}}, \bibinfo {author} {\bibfnamefont {J.}~\bibnamefont {Wosnitza}}, \bibinfo {author} {\bibfnamefont {C.~V.}\ \bibnamefont {Topping}}, \bibinfo {author} {\bibfnamefont {K.~E.}\ \bibnamefont {Carreiro}}, \bibinfo {author} {\bibfnamefont {H.~E.}\ \bibnamefont {Tran}}, \bibinfo {author} {\bibfnamefont {Z.~E.}\ \bibnamefont {Manson}}, \ and\ \bibinfo {author} {\bibfnamefont {J.~L.}\ \bibnamefont {Manson}},\ }\href {\doibase 10.1103/PhysRevB.95.024404} {\bibfield  {journal} {\bibinfo  {journal} {Phys. Rev. B}\ }\textbf {\bibinfo {volume} {95}},\ \bibinfo {pages} {024404} (\bibinfo {year} {2017})}\BibitemShut {NoStop}%
\bibitem [{\citenamefont {Williams}\ \emph {et~al.}(2020)\citenamefont {Williams}, \citenamefont {Blackmore}, \citenamefont {Curley}, \citenamefont {Lees}, \citenamefont {Birnbaum}, \citenamefont {Singleton}, \citenamefont {Huddart}, \citenamefont {Hicken}, \citenamefont {Lancaster}, \citenamefont {Blundell}, \citenamefont {Xiao}, \citenamefont {Ozarowski}, \citenamefont {Pratt}, \citenamefont {Voneshen}, \citenamefont {Guguchia}, \citenamefont {Baines}, \citenamefont {Schlueter}, \citenamefont {Villa}, \citenamefont {Manson},\ and\ \citenamefont {Goddard}}]{Willaims_2020_haldane}%
  \BibitemOpen
  \bibfield  {author} {\bibinfo {author} {\bibfnamefont {R.~C.}\ \bibnamefont {Williams}}, \bibinfo {author} {\bibfnamefont {W.~J.~A.}\ \bibnamefont {Blackmore}}, \bibinfo {author} {\bibfnamefont {S.~P.~M.}\ \bibnamefont {Curley}}, \bibinfo {author} {\bibfnamefont {M.~R.}\ \bibnamefont {Lees}}, \bibinfo {author} {\bibfnamefont {S.~M.}\ \bibnamefont {Birnbaum}}, \bibinfo {author} {\bibfnamefont {J.}~\bibnamefont {Singleton}}, \bibinfo {author} {\bibfnamefont {B.~M.}\ \bibnamefont {Huddart}}, \bibinfo {author} {\bibfnamefont {T.~J.}\ \bibnamefont {Hicken}}, \bibinfo {author} {\bibfnamefont {T.}~\bibnamefont {Lancaster}}, \bibinfo {author} {\bibfnamefont {S.~J.}\ \bibnamefont {Blundell}}, \bibinfo {author} {\bibfnamefont {F.}~\bibnamefont {Xiao}}, \bibinfo {author} {\bibfnamefont {A.}~\bibnamefont {Ozarowski}}, \bibinfo {author} {\bibfnamefont {F.~L.}\ \bibnamefont {Pratt}}, \bibinfo {author} {\bibfnamefont {D.~J.}\ \bibnamefont {Voneshen}}, \bibinfo {author} {\bibfnamefont {Z.}~\bibnamefont {Guguchia}},
  \bibinfo {author} {\bibfnamefont {C.}~\bibnamefont {Baines}}, \bibinfo {author} {\bibfnamefont {J.~A.}\ \bibnamefont {Schlueter}}, \bibinfo {author} {\bibfnamefont {D.~Y.}\ \bibnamefont {Villa}}, \bibinfo {author} {\bibfnamefont {J.~L.}\ \bibnamefont {Manson}}, \ and\ \bibinfo {author} {\bibfnamefont {P.~A.}\ \bibnamefont {Goddard}},\ }\href {\doibase 10.1103/PhysRevResearch.2.013082} {\bibfield  {journal} {\bibinfo  {journal} {Phys. Rev. Res.}\ }\textbf {\bibinfo {volume} {2}},\ \bibinfo {pages} {013082} (\bibinfo {year} {2020})}\BibitemShut {NoStop}%
\bibitem [{\citenamefont {Thorarinsdottir}\ and\ \citenamefont {Harris}(2020)}]{Thorarinsdottir_MOM_rev_2020}%
  \BibitemOpen
  \bibfield  {author} {\bibinfo {author} {\bibfnamefont {A.~E.}\ \bibnamefont {Thorarinsdottir}}\ and\ \bibinfo {author} {\bibfnamefont {T.~D.}\ \bibnamefont {Harris}},\ }\href {https://pubs.acs.org/doi/10.1021/acs.chemrev.9b00666} {\bibfield  {journal} {\bibinfo  {journal} {Chemical reviews}\ }\textbf {\bibinfo {volume} {120}},\ \bibinfo {pages} {8716} (\bibinfo {year} {2020})}\BibitemShut {NoStop}%
\bibitem [{\citenamefont {Coronado}(2020)}]{Coronado_2020_molecular}%
  \BibitemOpen
  \bibfield  {author} {\bibinfo {author} {\bibfnamefont {E.}~\bibnamefont {Coronado}},\ }\href {https://www.nature.com/articles/s41578-019-0146-8} {\bibfield  {journal} {\bibinfo  {journal} {Nature Reviews Materials}\ }\textbf {\bibinfo {volume} {5}},\ \bibinfo {pages} {87} (\bibinfo {year} {2020})}\BibitemShut {NoStop}%
\bibitem [{\citenamefont {Chilton}(2022)}]{Chilton_2022_molecular}%
  \BibitemOpen
  \bibfield  {author} {\bibinfo {author} {\bibfnamefont {N.~F.}\ \bibnamefont {Chilton}},\ }\href {https://doi.org/10.1146/annurev-matsci-081420-042553} {\bibfield  {journal} {\bibinfo  {journal} {Annual Review of Materials Research}\ }\textbf {\bibinfo {volume} {52}},\ \bibinfo {pages} {79} (\bibinfo {year} {2022})}\BibitemShut {NoStop}%
\bibitem [{\citenamefont {Goddard}\ \emph {et~al.}(2012)\citenamefont {Goddard}, \citenamefont {Manson}, \citenamefont {Singleton}, \citenamefont {Franke}, \citenamefont {Lancaster}, \citenamefont {Steele}, \citenamefont {Blundell}, \citenamefont {Baines}, \citenamefont {Pratt}, \citenamefont {McDonald}, \citenamefont {Ayala-Valenzuela}, \citenamefont {Corbey}, \citenamefont {Southerland}, \citenamefont {Sengupta},\ and\ \citenamefont {Schlueter}}]{Goddard_dimensional_2012}%
  \BibitemOpen
  \bibfield  {author} {\bibinfo {author} {\bibfnamefont {P.~A.}\ \bibnamefont {Goddard}}, \bibinfo {author} {\bibfnamefont {J.~L.}\ \bibnamefont {Manson}}, \bibinfo {author} {\bibfnamefont {J.}~\bibnamefont {Singleton}}, \bibinfo {author} {\bibfnamefont {I.}~\bibnamefont {Franke}}, \bibinfo {author} {\bibfnamefont {T.}~\bibnamefont {Lancaster}}, \bibinfo {author} {\bibfnamefont {A.~J.}\ \bibnamefont {Steele}}, \bibinfo {author} {\bibfnamefont {S.~J.}\ \bibnamefont {Blundell}}, \bibinfo {author} {\bibfnamefont {C.}~\bibnamefont {Baines}}, \bibinfo {author} {\bibfnamefont {F.~L.}\ \bibnamefont {Pratt}}, \bibinfo {author} {\bibfnamefont {R.~D.}\ \bibnamefont {McDonald}}, \bibinfo {author} {\bibfnamefont {O.~E.}\ \bibnamefont {Ayala-Valenzuela}}, \bibinfo {author} {\bibfnamefont {J.~F.}\ \bibnamefont {Corbey}}, \bibinfo {author} {\bibfnamefont {H.~I.}\ \bibnamefont {Southerland}}, \bibinfo {author} {\bibfnamefont {P.}~\bibnamefont {Sengupta}}, \ and\ \bibinfo {author} {\bibfnamefont {J.~A.}\ \bibnamefont
  {Schlueter}},\ }\href {\doibase 10.1103/PhysRevLett.108.077208} {\bibfield  {journal} {\bibinfo  {journal} {Phys. Rev. Lett.}\ }\textbf {\bibinfo {volume} {108}},\ \bibinfo {pages} {077208} (\bibinfo {year} {2012})}\BibitemShut {NoStop}%
\bibitem [{\citenamefont {Pitcairn}\ \emph {et~al.}(2023)\citenamefont {Pitcairn}, \citenamefont {Iliceto}, \citenamefont {Ca{\~n}adillas-Delgado}, \citenamefont {Fabelo}, \citenamefont {Liu}, \citenamefont {Balz}, \citenamefont {Weilhard}, \citenamefont {Argent}, \citenamefont {Morris},\ and\ \citenamefont {Cliffe}}]{Pitcarin_2023}%
  \BibitemOpen
  \bibfield  {author} {\bibinfo {author} {\bibfnamefont {J.}~\bibnamefont {Pitcairn}}, \bibinfo {author} {\bibfnamefont {A.}~\bibnamefont {Iliceto}}, \bibinfo {author} {\bibfnamefont {L.}~\bibnamefont {Ca{\~n}adillas-Delgado}}, \bibinfo {author} {\bibfnamefont {O.}~\bibnamefont {Fabelo}}, \bibinfo {author} {\bibfnamefont {C.}~\bibnamefont {Liu}}, \bibinfo {author} {\bibfnamefont {C.}~\bibnamefont {Balz}}, \bibinfo {author} {\bibfnamefont {A.}~\bibnamefont {Weilhard}}, \bibinfo {author} {\bibfnamefont {S.~P.}\ \bibnamefont {Argent}}, \bibinfo {author} {\bibfnamefont {A.~J.}\ \bibnamefont {Morris}}, \ and\ \bibinfo {author} {\bibfnamefont {M.~J.}\ \bibnamefont {Cliffe}},\ }\href {https://pubs.acs.org/doi/full/10.1021/jacs.2c10916} {\bibfield  {journal} {\bibinfo  {journal} {Journal of the American Chemical Society}\ }\textbf {\bibinfo {volume} {145}},\ \bibinfo {pages} {1783} (\bibinfo {year} {2023})}\BibitemShut {NoStop}%
\bibitem [{\citenamefont {Wang}\ \emph {et~al.}(2024)\citenamefont {Wang}, \citenamefont {Fu}, \citenamefont {Takatsu}, \citenamefont {Tassel}, \citenamefont {Hayashi}, \citenamefont {Cao}, \citenamefont {Bataille}, \citenamefont {Koo}, \citenamefont {Ouyang}, \citenamefont {Whangbo} \emph {et~al.}}]{Wang_dimensional_2024}%
  \BibitemOpen
  \bibfield  {author} {\bibinfo {author} {\bibfnamefont {Y.}~\bibnamefont {Wang}}, \bibinfo {author} {\bibfnamefont {P.}~\bibnamefont {Fu}}, \bibinfo {author} {\bibfnamefont {H.}~\bibnamefont {Takatsu}}, \bibinfo {author} {\bibfnamefont {C.}~\bibnamefont {Tassel}}, \bibinfo {author} {\bibfnamefont {N.}~\bibnamefont {Hayashi}}, \bibinfo {author} {\bibfnamefont {J.}~\bibnamefont {Cao}}, \bibinfo {author} {\bibfnamefont {T.}~\bibnamefont {Bataille}}, \bibinfo {author} {\bibfnamefont {H.-J.}\ \bibnamefont {Koo}}, \bibinfo {author} {\bibfnamefont {Z.}~\bibnamefont {Ouyang}}, \bibinfo {author} {\bibfnamefont {M.-H.}\ \bibnamefont {Whangbo}},  \emph {et~al.},\ }\href {https://doi.org/10.1021/jacs.3c13902} {\bibfield  {journal} {\bibinfo  {journal} {Journal of the American Chemical Society}\ } (\bibinfo {year} {2024})}\BibitemShut {NoStop}%
\bibitem [{\citenamefont {Schlueter}\ \emph {et~al.}(2012)\citenamefont {Schlueter}, \citenamefont {Park}, \citenamefont {Halder}, \citenamefont {Armand}, \citenamefont {Dunmars}, \citenamefont {Chapman}, \citenamefont {Manson}, \citenamefont {Singleton}, \citenamefont {McDonald}, \citenamefont {Plonczak} \emph {et~al.}}]{Schlueter_2012}%
  \BibitemOpen
  \bibfield  {author} {\bibinfo {author} {\bibfnamefont {J.~A.}\ \bibnamefont {Schlueter}}, \bibinfo {author} {\bibfnamefont {H.}~\bibnamefont {Park}}, \bibinfo {author} {\bibfnamefont {G.~J.}\ \bibnamefont {Halder}}, \bibinfo {author} {\bibfnamefont {W.~R.}\ \bibnamefont {Armand}}, \bibinfo {author} {\bibfnamefont {C.}~\bibnamefont {Dunmars}}, \bibinfo {author} {\bibfnamefont {K.~W.}\ \bibnamefont {Chapman}}, \bibinfo {author} {\bibfnamefont {J.~L.}\ \bibnamefont {Manson}}, \bibinfo {author} {\bibfnamefont {J.}~\bibnamefont {Singleton}}, \bibinfo {author} {\bibfnamefont {R.}~\bibnamefont {McDonald}}, \bibinfo {author} {\bibfnamefont {A.}~\bibnamefont {Plonczak}},  \emph {et~al.},\ }\href {https://0-doi-org.pugwash.lib.warwick.ac.uk/10.1021/ic201924q} {\bibfield  {journal} {\bibinfo  {journal} {Inorganic chemistry}\ }\textbf {\bibinfo {volume} {51}},\ \bibinfo {pages} {2121} (\bibinfo {year} {2012})}\BibitemShut {NoStop}%
\bibitem [{\citenamefont {Cortijo}\ \emph {et~al.}(2013)\citenamefont {Cortijo}, \citenamefont {Herrero}, \citenamefont {Jimenez-Aparicio},\ and\ \citenamefont {Matesanz}}]{Cortijo_2013}%
  \BibitemOpen
  \bibfield  {author} {\bibinfo {author} {\bibfnamefont {M.}~\bibnamefont {Cortijo}}, \bibinfo {author} {\bibfnamefont {S.}~\bibnamefont {Herrero}}, \bibinfo {author} {\bibfnamefont {R.}~\bibnamefont {Jimenez-Aparicio}}, \ and\ \bibinfo {author} {\bibfnamefont {E.}~\bibnamefont {Matesanz}},\ }\href {https://doi.org/10.1021/ic400632c} {\bibfield  {journal} {\bibinfo  {journal} {Inorganic chemistry}\ }\textbf {\bibinfo {volume} {52}},\ \bibinfo {pages} {7087} (\bibinfo {year} {2013})}\BibitemShut {NoStop}%
\bibitem [{\citenamefont {Cortijo}\ \emph {et~al.}(2014)\citenamefont {Cortijo}, \citenamefont {Herrero}, \citenamefont {Jimenez-Aparicio}, \citenamefont {Perles}, \citenamefont {Priego},\ and\ \citenamefont {Torroba}}]{Cortijo_2014}%
  \BibitemOpen
  \bibfield  {author} {\bibinfo {author} {\bibfnamefont {M.}~\bibnamefont {Cortijo}}, \bibinfo {author} {\bibfnamefont {S.}~\bibnamefont {Herrero}}, \bibinfo {author} {\bibfnamefont {R.}~\bibnamefont {Jimenez-Aparicio}}, \bibinfo {author} {\bibfnamefont {J.}~\bibnamefont {Perles}}, \bibinfo {author} {\bibfnamefont {J.~L.}\ \bibnamefont {Priego}}, \ and\ \bibinfo {author} {\bibfnamefont {J.}~\bibnamefont {Torroba}},\ }\href {https://doi.org/10.1021/cg401590w} {\bibfield  {journal} {\bibinfo  {journal} {Crystal growth \& design}\ }\textbf {\bibinfo {volume} {14}},\ \bibinfo {pages} {716} (\bibinfo {year} {2014})}\BibitemShut {NoStop}%
\bibitem [{\citenamefont {Liu}\ \emph {et~al.}(2016)\citenamefont {Liu}, \citenamefont {Goddard}, \citenamefont {Singleton}, \citenamefont {Brambleby}, \citenamefont {Foronda}, \citenamefont {M\"{o}ller}, \citenamefont {Kohama}, \citenamefont {Ghannadzadeh}, \citenamefont {Ardavan}, \citenamefont {Blundell} \emph {et~al.}}]{Liu_2016_correlations}%
  \BibitemOpen
  \bibfield  {author} {\bibinfo {author} {\bibfnamefont {J.}~\bibnamefont {Liu}}, \bibinfo {author} {\bibfnamefont {P.~A.}\ \bibnamefont {Goddard}}, \bibinfo {author} {\bibfnamefont {J.}~\bibnamefont {Singleton}}, \bibinfo {author} {\bibfnamefont {J.}~\bibnamefont {Brambleby}}, \bibinfo {author} {\bibfnamefont {F.}~\bibnamefont {Foronda}}, \bibinfo {author} {\bibfnamefont {J.~S.}\ \bibnamefont {M\"{o}ller}}, \bibinfo {author} {\bibfnamefont {Y.}~\bibnamefont {Kohama}}, \bibinfo {author} {\bibfnamefont {S.}~\bibnamefont {Ghannadzadeh}}, \bibinfo {author} {\bibfnamefont {A.}~\bibnamefont {Ardavan}}, \bibinfo {author} {\bibfnamefont {S.~J.}\ \bibnamefont {Blundell}},  \emph {et~al.},\ }\href {https://doi.org/10.1021/acs.inorgchem.5b02991} {\bibfield  {journal} {\bibinfo  {journal} {Inorganic Chemistry}\ }\textbf {\bibinfo {volume} {55}},\ \bibinfo {pages} {3515} (\bibinfo {year} {2016})}\BibitemShut {NoStop}%
\bibitem [{\citenamefont {Kubus}\ \emph {et~al.}(2018)\citenamefont {Kubus}, \citenamefont {Lanza}, \citenamefont {Scatena}, \citenamefont {Dos~Santos}, \citenamefont {Wehinger}, \citenamefont {Casati}, \citenamefont {Fiolka}, \citenamefont {Keller}, \citenamefont {Macchi}, \citenamefont {Ruegg} \emph {et~al.}}]{Kubus_2018}%
  \BibitemOpen
  \bibfield  {author} {\bibinfo {author} {\bibfnamefont {M.}~\bibnamefont {Kubus}}, \bibinfo {author} {\bibfnamefont {A.}~\bibnamefont {Lanza}}, \bibinfo {author} {\bibfnamefont {R.}~\bibnamefont {Scatena}}, \bibinfo {author} {\bibfnamefont {L.~H.}\ \bibnamefont {Dos~Santos}}, \bibinfo {author} {\bibfnamefont {B.}~\bibnamefont {Wehinger}}, \bibinfo {author} {\bibfnamefont {N.}~\bibnamefont {Casati}}, \bibinfo {author} {\bibfnamefont {C.}~\bibnamefont {Fiolka}}, \bibinfo {author} {\bibfnamefont {L.}~\bibnamefont {Keller}}, \bibinfo {author} {\bibfnamefont {P.}~\bibnamefont {Macchi}}, \bibinfo {author} {\bibfnamefont {C.}~\bibnamefont {Ruegg}},  \emph {et~al.},\ }\href {https://doi.org/10.1021/acs.inorgchem.7b03150} {\bibfield  {journal} {\bibinfo  {journal} {Inorganic chemistry}\ }\textbf {\bibinfo {volume} {57}},\ \bibinfo {pages} {4934} (\bibinfo {year} {2018})}\BibitemShut {NoStop}%
\bibitem [{\citenamefont {Blackmore}\ \emph {et~al.}(2022)\citenamefont {Blackmore}, \citenamefont {Curley}, \citenamefont {Williams}, \citenamefont {Vaidya}, \citenamefont {Singleton}, \citenamefont {Birnbaum}, \citenamefont {Ozarowski}, \citenamefont {Schlueter}, \citenamefont {Chen}, \citenamefont {Gillon}, \citenamefont {Goukassov}, \citenamefont {Kibalin}, \citenamefont {Villa}, \citenamefont {Villa}, \citenamefont {Manson},\ and\ \citenamefont {Goddard}}]{Blackmore-correlations}%
  \BibitemOpen
  \bibfield  {author} {\bibinfo {author} {\bibfnamefont {W.~J.~A.}\ \bibnamefont {Blackmore}}, \bibinfo {author} {\bibfnamefont {S.~P.~M.}\ \bibnamefont {Curley}}, \bibinfo {author} {\bibfnamefont {R.~C.}\ \bibnamefont {Williams}}, \bibinfo {author} {\bibfnamefont {S.}~\bibnamefont {Vaidya}}, \bibinfo {author} {\bibfnamefont {J.}~\bibnamefont {Singleton}}, \bibinfo {author} {\bibfnamefont {S.}~\bibnamefont {Birnbaum}}, \bibinfo {author} {\bibfnamefont {A.}~\bibnamefont {Ozarowski}}, \bibinfo {author} {\bibfnamefont {J.~A.}\ \bibnamefont {Schlueter}}, \bibinfo {author} {\bibfnamefont {Y.-S.}\ \bibnamefont {Chen}}, \bibinfo {author} {\bibfnamefont {B.}~\bibnamefont {Gillon}}, \bibinfo {author} {\bibfnamefont {A.}~\bibnamefont {Goukassov}}, \bibinfo {author} {\bibfnamefont {I.}~\bibnamefont {Kibalin}}, \bibinfo {author} {\bibfnamefont {D.~Y.}\ \bibnamefont {Villa}}, \bibinfo {author} {\bibfnamefont {J.~A.}\ \bibnamefont {Villa}}, \bibinfo {author} {\bibfnamefont {J.~L.}\ \bibnamefont {Manson}}, \ and\ \bibinfo
  {author} {\bibfnamefont {P.~A.}\ \bibnamefont {Goddard}},\ }\href {\doibase 10.1021/acs.inorgchem.1c02483} {\bibfield  {journal} {\bibinfo  {journal} {Inorganic Chemistry}\ }\textbf {\bibinfo {volume} {61}},\ \bibinfo {pages} {141} (\bibinfo {year} {2022})}\BibitemShut {NoStop}%
\bibitem [{\citenamefont {Geers}\ \emph {et~al.}(2023{\natexlab{a}})\citenamefont {Geers}, \citenamefont {Lee}, \citenamefont {Ling}, \citenamefont {Fabelo}, \citenamefont {Ca{\~n}adillas-Delgado},\ and\ \citenamefont {Cliffe}}]{Geers_2023_NCS}%
  \BibitemOpen
  \bibfield  {author} {\bibinfo {author} {\bibfnamefont {M.}~\bibnamefont {Geers}}, \bibinfo {author} {\bibfnamefont {J.~Y.}\ \bibnamefont {Lee}}, \bibinfo {author} {\bibfnamefont {S.}~\bibnamefont {Ling}}, \bibinfo {author} {\bibfnamefont {O.}~\bibnamefont {Fabelo}}, \bibinfo {author} {\bibfnamefont {L.}~\bibnamefont {Ca{\~n}adillas-Delgado}}, \ and\ \bibinfo {author} {\bibfnamefont {M.~J.}\ \bibnamefont {Cliffe}},\ }\href {https://doi.org/10.1039/D2SC06861C} {\bibfield  {journal} {\bibinfo  {journal} {Chemical Science}\ }\textbf {\bibinfo {volume} {14}},\ \bibinfo {pages} {3531} (\bibinfo {year} {2023}{\natexlab{a}})}\BibitemShut {NoStop}%
\bibitem [{\citenamefont {Woodward}\ \emph {et~al.}(2007)\citenamefont {Woodward}, \citenamefont {Gibson}, \citenamefont {Jameson}, \citenamefont {Landee}, \citenamefont {Turnbull},\ and\ \citenamefont {Willett}}]{Woodward_counter-ion_2007}%
  \BibitemOpen
  \bibfield  {author} {\bibinfo {author} {\bibfnamefont {F.~M.}\ \bibnamefont {Woodward}}, \bibinfo {author} {\bibfnamefont {P.~J.}\ \bibnamefont {Gibson}}, \bibinfo {author} {\bibfnamefont {G.~B.}\ \bibnamefont {Jameson}}, \bibinfo {author} {\bibfnamefont {C.~P.}\ \bibnamefont {Landee}}, \bibinfo {author} {\bibfnamefont {M.~M.}\ \bibnamefont {Turnbull}}, \ and\ \bibinfo {author} {\bibfnamefont {R.~D.}\ \bibnamefont {Willett}},\ }\href {https://pubs.acs.org/doi/10.1021/ic0621392} {\bibfield  {journal} {\bibinfo  {journal} {Inorganic chemistry}\ }\textbf {\bibinfo {volume} {46}},\ \bibinfo {pages} {4256} (\bibinfo {year} {2007})}\BibitemShut {NoStop}%
\bibitem [{\citenamefont {Manson}\ \emph {et~al.}(2011)\citenamefont {Manson}, \citenamefont {Lapidus}, \citenamefont {Stephens}, \citenamefont {Peterson}, \citenamefont {Carreiro}, \citenamefont {Southerland}, \citenamefont {Lancaster}, \citenamefont {Blundell}, \citenamefont {Steele}, \citenamefont {Goddard} \emph {et~al.}}]{Manson_counter-ion_2021}%
  \BibitemOpen
  \bibfield  {author} {\bibinfo {author} {\bibfnamefont {J.~L.}\ \bibnamefont {Manson}}, \bibinfo {author} {\bibfnamefont {S.~H.}\ \bibnamefont {Lapidus}}, \bibinfo {author} {\bibfnamefont {P.~W.}\ \bibnamefont {Stephens}}, \bibinfo {author} {\bibfnamefont {P.~K.}\ \bibnamefont {Peterson}}, \bibinfo {author} {\bibfnamefont {K.~E.}\ \bibnamefont {Carreiro}}, \bibinfo {author} {\bibfnamefont {H.~I.}\ \bibnamefont {Southerland}}, \bibinfo {author} {\bibfnamefont {T.}~\bibnamefont {Lancaster}}, \bibinfo {author} {\bibfnamefont {S.~J.}\ \bibnamefont {Blundell}}, \bibinfo {author} {\bibfnamefont {A.~J.}\ \bibnamefont {Steele}}, \bibinfo {author} {\bibfnamefont {P.~A.}\ \bibnamefont {Goddard}},  \emph {et~al.},\ }\href {https://doi.org/10.1021/ic102532h} {\bibfield  {journal} {\bibinfo  {journal} {Inorganic chemistry}\ }\textbf {\bibinfo {volume} {50}},\ \bibinfo {pages} {5990} (\bibinfo {year} {2011})}\BibitemShut {NoStop}%
\bibitem [{\citenamefont {Pajerowski}\ \emph {et~al.}(2022)\citenamefont {Pajerowski}, \citenamefont {Podlesnyak}, \citenamefont {Herbrych},\ and\ \citenamefont {Manson}}]{Pajerowski_2022}%
  \BibitemOpen
  \bibfield  {author} {\bibinfo {author} {\bibfnamefont {D.~M.}\ \bibnamefont {Pajerowski}}, \bibinfo {author} {\bibfnamefont {A.~P.}\ \bibnamefont {Podlesnyak}}, \bibinfo {author} {\bibfnamefont {J.}~\bibnamefont {Herbrych}}, \ and\ \bibinfo {author} {\bibfnamefont {J.}~\bibnamefont {Manson}},\ }\href {\doibase 10.1103/PhysRevB.105.134420} {\bibfield  {journal} {\bibinfo  {journal} {Phys. Rev. B}\ }\textbf {\bibinfo {volume} {105}},\ \bibinfo {pages} {134420} (\bibinfo {year} {2022})}\BibitemShut {NoStop}%
\bibitem [{\citenamefont {Coak}\ \emph {et~al.}(2023)\citenamefont {Coak}, \citenamefont {Curley}, \citenamefont {Hawkhead}, \citenamefont {Tidey}, \citenamefont {Graf}, \citenamefont {Clark}, \citenamefont {Sengupta}, \citenamefont {Manson}, \citenamefont {Lancaster}, \citenamefont {Goddard},\ and\ \citenamefont {Manson}}]{Coak_2023}%
  \BibitemOpen
  \bibfield  {author} {\bibinfo {author} {\bibfnamefont {M.~J.}\ \bibnamefont {Coak}}, \bibinfo {author} {\bibfnamefont {S.~P.~M.}\ \bibnamefont {Curley}}, \bibinfo {author} {\bibfnamefont {Z.}~\bibnamefont {Hawkhead}}, \bibinfo {author} {\bibfnamefont {J.~P.}\ \bibnamefont {Tidey}}, \bibinfo {author} {\bibfnamefont {D.}~\bibnamefont {Graf}}, \bibinfo {author} {\bibfnamefont {S.~J.}\ \bibnamefont {Clark}}, \bibinfo {author} {\bibfnamefont {P.}~\bibnamefont {Sengupta}}, \bibinfo {author} {\bibfnamefont {Z.~E.}\ \bibnamefont {Manson}}, \bibinfo {author} {\bibfnamefont {T.}~\bibnamefont {Lancaster}}, \bibinfo {author} {\bibfnamefont {P.~A.}\ \bibnamefont {Goddard}}, \ and\ \bibinfo {author} {\bibfnamefont {J.~L.}\ \bibnamefont {Manson}},\ }\href {\doibase 10.1103/PhysRevB.108.224431} {\bibfield  {journal} {\bibinfo  {journal} {Phys. Rev. B}\ }\textbf {\bibinfo {volume} {108}},\ \bibinfo {pages} {224431} (\bibinfo {year} {2023})}\BibitemShut {NoStop}%
\bibitem [{\citenamefont {Geers}\ \emph {et~al.}(2023{\natexlab{b}})\citenamefont {Geers}, \citenamefont {Jarvis}, \citenamefont {Liu}, \citenamefont {Saxena}, \citenamefont {Pitcairn}, \citenamefont {Myatt}, \citenamefont {Hallweger}, \citenamefont {Kronawitter}, \citenamefont {Kieslich}, \citenamefont {Ling}, \citenamefont {Cairns}, \citenamefont {Daisenberger}, \citenamefont {Fabelo}, \citenamefont {Ca\~nadillas Delgado},\ and\ \citenamefont {Cliffe}}]{Geers_preesure_2023}%
  \BibitemOpen
  \bibfield  {author} {\bibinfo {author} {\bibfnamefont {M.}~\bibnamefont {Geers}}, \bibinfo {author} {\bibfnamefont {D.~M.}\ \bibnamefont {Jarvis}}, \bibinfo {author} {\bibfnamefont {C.}~\bibnamefont {Liu}}, \bibinfo {author} {\bibfnamefont {S.~S.}\ \bibnamefont {Saxena}}, \bibinfo {author} {\bibfnamefont {J.}~\bibnamefont {Pitcairn}}, \bibinfo {author} {\bibfnamefont {E.}~\bibnamefont {Myatt}}, \bibinfo {author} {\bibfnamefont {S.~A.}\ \bibnamefont {Hallweger}}, \bibinfo {author} {\bibfnamefont {S.~M.}\ \bibnamefont {Kronawitter}}, \bibinfo {author} {\bibfnamefont {G.}~\bibnamefont {Kieslich}}, \bibinfo {author} {\bibfnamefont {S.}~\bibnamefont {Ling}}, \bibinfo {author} {\bibfnamefont {A.~B.}\ \bibnamefont {Cairns}}, \bibinfo {author} {\bibfnamefont {D.}~\bibnamefont {Daisenberger}}, \bibinfo {author} {\bibfnamefont {O.}~\bibnamefont {Fabelo}}, \bibinfo {author} {\bibfnamefont {L.}~\bibnamefont {Ca\~nadillas Delgado}}, \ and\ \bibinfo {author} {\bibfnamefont {M.~J.}\ \bibnamefont {Cliffe}},\ }\href
  {\doibase 10.1103/PhysRevB.108.144439} {\bibfield  {journal} {\bibinfo  {journal} {Phys. Rev. B}\ }\textbf {\bibinfo {volume} {108}},\ \bibinfo {pages} {144439} (\bibinfo {year} {2023}{\natexlab{b}})}\BibitemShut {NoStop}%
\bibitem [{\citenamefont {Povarov}\ \emph {et~al.}(2024)\citenamefont {Povarov}, \citenamefont {Graf}, \citenamefont {Hauspurg}, \citenamefont {Zherlitsyn}, \citenamefont {Wosnitza}, \citenamefont {Sakurai}, \citenamefont {Ohta}, \citenamefont {Kimura}, \citenamefont {Nojiri}, \citenamefont {Garlea} \emph {et~al.}}]{Povarov_2024}%
  \BibitemOpen
  \bibfield  {author} {\bibinfo {author} {\bibfnamefont {K.~Y.}\ \bibnamefont {Povarov}}, \bibinfo {author} {\bibfnamefont {D.~E.}\ \bibnamefont {Graf}}, \bibinfo {author} {\bibfnamefont {A.}~\bibnamefont {Hauspurg}}, \bibinfo {author} {\bibfnamefont {S.}~\bibnamefont {Zherlitsyn}}, \bibinfo {author} {\bibfnamefont {J.}~\bibnamefont {Wosnitza}}, \bibinfo {author} {\bibfnamefont {T.}~\bibnamefont {Sakurai}}, \bibinfo {author} {\bibfnamefont {H.}~\bibnamefont {Ohta}}, \bibinfo {author} {\bibfnamefont {S.}~\bibnamefont {Kimura}}, \bibinfo {author} {\bibfnamefont {H.}~\bibnamefont {Nojiri}}, \bibinfo {author} {\bibfnamefont {V.~O.}\ \bibnamefont {Garlea}},  \emph {et~al.},\ }\href {https://doi.org/10.1038/s41467-024-46527-x} {\bibfield  {journal} {\bibinfo  {journal} {Nature Communications}\ }\textbf {\bibinfo {volume} {15}},\ \bibinfo {pages} {2295} (\bibinfo {year} {2024})}\BibitemShut {NoStop}%
\bibitem [{\citenamefont {Feyerherm}\ \emph {et~al.}(2000)\citenamefont {Feyerherm}, \citenamefont {Abens}, \citenamefont {G{\"u}nther}, \citenamefont {Ishida}, \citenamefont {Mei{\ss}ner}, \citenamefont {Meschke}, \citenamefont {Nogami},\ and\ \citenamefont {Steiner}}]{Feyerherm_stag_2000}%
  \BibitemOpen
  \bibfield  {author} {\bibinfo {author} {\bibfnamefont {R.}~\bibnamefont {Feyerherm}}, \bibinfo {author} {\bibfnamefont {S.}~\bibnamefont {Abens}}, \bibinfo {author} {\bibfnamefont {D.}~\bibnamefont {G{\"u}nther}}, \bibinfo {author} {\bibfnamefont {T.}~\bibnamefont {Ishida}}, \bibinfo {author} {\bibfnamefont {M.}~\bibnamefont {Mei{\ss}ner}}, \bibinfo {author} {\bibfnamefont {M.}~\bibnamefont {Meschke}}, \bibinfo {author} {\bibfnamefont {T.}~\bibnamefont {Nogami}}, \ and\ \bibinfo {author} {\bibfnamefont {M.}~\bibnamefont {Steiner}},\ }\href {https://iopscience.iop.org/article/10.1088/0953-8984/12/39/312} {\bibfield  {journal} {\bibinfo  {journal} {Journal of Physics: Condensed Matter}\ }\textbf {\bibinfo {volume} {12}},\ \bibinfo {pages} {8495} (\bibinfo {year} {2000})}\BibitemShut {NoStop}%
\bibitem [{\citenamefont {Zvyagin}\ \emph {et~al.}(2004)\citenamefont {Zvyagin}, \citenamefont {Kolezhuk}, \citenamefont {Krzystek},\ and\ \citenamefont {Feyerherm}}]{Zvyagin_SG_2004}%
  \BibitemOpen
  \bibfield  {author} {\bibinfo {author} {\bibfnamefont {S.~A.}\ \bibnamefont {Zvyagin}}, \bibinfo {author} {\bibfnamefont {A.~K.}\ \bibnamefont {Kolezhuk}}, \bibinfo {author} {\bibfnamefont {J.}~\bibnamefont {Krzystek}}, \ and\ \bibinfo {author} {\bibfnamefont {R.}~\bibnamefont {Feyerherm}},\ }\href {\doibase 10.1103/PhysRevLett.93.027201} {\bibfield  {journal} {\bibinfo  {journal} {Phys. Rev. Lett.}\ }\textbf {\bibinfo {volume} {93}},\ \bibinfo {pages} {027201} (\bibinfo {year} {2004})}\BibitemShut {NoStop}%
\bibitem [{\citenamefont {Liu}\ \emph {et~al.}(2019)\citenamefont {Liu}, \citenamefont {Kittaka}, \citenamefont {Johnson}, \citenamefont {Lancaster}, \citenamefont {Singleton}, \citenamefont {Sakakibara}, \citenamefont {Kohama}, \citenamefont {van Tol}, \citenamefont {Ardavan}, \citenamefont {Williams}, \citenamefont {Blundell}, \citenamefont {Manson}, \citenamefont {Manson},\ and\ \citenamefont {Goddard}}]{Liu_chiral_2019}%
  \BibitemOpen
  \bibfield  {author} {\bibinfo {author} {\bibfnamefont {J.}~\bibnamefont {Liu}}, \bibinfo {author} {\bibfnamefont {S.}~\bibnamefont {Kittaka}}, \bibinfo {author} {\bibfnamefont {R.~D.}\ \bibnamefont {Johnson}}, \bibinfo {author} {\bibfnamefont {T.}~\bibnamefont {Lancaster}}, \bibinfo {author} {\bibfnamefont {J.}~\bibnamefont {Singleton}}, \bibinfo {author} {\bibfnamefont {T.}~\bibnamefont {Sakakibara}}, \bibinfo {author} {\bibfnamefont {Y.}~\bibnamefont {Kohama}}, \bibinfo {author} {\bibfnamefont {J.}~\bibnamefont {van Tol}}, \bibinfo {author} {\bibfnamefont {A.}~\bibnamefont {Ardavan}}, \bibinfo {author} {\bibfnamefont {B.~H.}\ \bibnamefont {Williams}}, \bibinfo {author} {\bibfnamefont {S.~J.}\ \bibnamefont {Blundell}}, \bibinfo {author} {\bibfnamefont {Z.~E.}\ \bibnamefont {Manson}}, \bibinfo {author} {\bibfnamefont {J.~L.}\ \bibnamefont {Manson}}, \ and\ \bibinfo {author} {\bibfnamefont {P.~A.}\ \bibnamefont {Goddard}},\ }\href {\doibase 10.1103/PhysRevLett.122.057207} {\bibfield  {journal} {\bibinfo
  {journal} {Phys. Rev. Lett.}\ }\textbf {\bibinfo {volume} {122}},\ \bibinfo {pages} {057207} (\bibinfo {year} {2019})}\BibitemShut {NoStop}%
\bibitem [{\citenamefont {Huddart}\ \emph {et~al.}(2021)\citenamefont {Huddart}, \citenamefont {Gomil\ifmmode~\check{s}\else \v{s}\fi{}ek}, \citenamefont {Hicken}, \citenamefont {Pratt}, \citenamefont {Blundell}, \citenamefont {Goddard}, \citenamefont {Kaech}, \citenamefont {Manson},\ and\ \citenamefont {Lancaster}}]{Huddart_spin_transport_2021}%
  \BibitemOpen
  \bibfield  {author} {\bibinfo {author} {\bibfnamefont {B.~M.}\ \bibnamefont {Huddart}}, \bibinfo {author} {\bibfnamefont {M.}~\bibnamefont {Gomil\ifmmode~\check{s}\else \v{s}\fi{}ek}}, \bibinfo {author} {\bibfnamefont {T.~J.}\ \bibnamefont {Hicken}}, \bibinfo {author} {\bibfnamefont {F.~L.}\ \bibnamefont {Pratt}}, \bibinfo {author} {\bibfnamefont {S.~J.}\ \bibnamefont {Blundell}}, \bibinfo {author} {\bibfnamefont {P.~A.}\ \bibnamefont {Goddard}}, \bibinfo {author} {\bibfnamefont {S.~J.}\ \bibnamefont {Kaech}}, \bibinfo {author} {\bibfnamefont {J.~L.}\ \bibnamefont {Manson}}, \ and\ \bibinfo {author} {\bibfnamefont {T.}~\bibnamefont {Lancaster}},\ }\href {\doibase 10.1103/PhysRevB.103.L060405} {\bibfield  {journal} {\bibinfo  {journal} {Phys. Rev. B}\ }\textbf {\bibinfo {volume} {103}},\ \bibinfo {pages} {L060405} (\bibinfo {year} {2021})}\BibitemShut {NoStop}%
\bibitem [{\citenamefont {Oshikawa}\ and\ \citenamefont {Affleck}(1997)}]{Oshikawa_SG_1997}%
  \BibitemOpen
  \bibfield  {author} {\bibinfo {author} {\bibfnamefont {M.}~\bibnamefont {Oshikawa}}\ and\ \bibinfo {author} {\bibfnamefont {I.}~\bibnamefont {Affleck}},\ }\href {\doibase 10.1103/PhysRevLett.79.2883} {\bibfield  {journal} {\bibinfo  {journal} {Phys. Rev. Lett.}\ }\textbf {\bibinfo {volume} {79}},\ \bibinfo {pages} {2883} (\bibinfo {year} {1997})}\BibitemShut {NoStop}%
\bibitem [{\citenamefont {Affleck}\ and\ \citenamefont {Oshikawa}(1999)}]{Affleck_SG_1999}%
  \BibitemOpen
  \bibfield  {author} {\bibinfo {author} {\bibfnamefont {I.}~\bibnamefont {Affleck}}\ and\ \bibinfo {author} {\bibfnamefont {M.}~\bibnamefont {Oshikawa}},\ }\href {\doibase 10.1103/PhysRevB.60.1038} {\bibfield  {journal} {\bibinfo  {journal} {Phys. Rev. B}\ }\textbf {\bibinfo {volume} {60}},\ \bibinfo {pages} {1038} (\bibinfo {year} {1999})}\BibitemShut {NoStop}%
\bibitem [{\citenamefont {Feyerherm}\ \emph {et~al.}(2004)\citenamefont {Feyerherm}, \citenamefont {Loose}, \citenamefont {Ishida}, \citenamefont {Nogami}, \citenamefont {Kreitlow}, \citenamefont {Baabe}, \citenamefont {Litterst}, \citenamefont {S\"ullow}, \citenamefont {Klauss},\ and\ \citenamefont {Doll}}]{FeCl2}%
  \BibitemOpen
  \bibfield  {author} {\bibinfo {author} {\bibfnamefont {R.}~\bibnamefont {Feyerherm}}, \bibinfo {author} {\bibfnamefont {A.}~\bibnamefont {Loose}}, \bibinfo {author} {\bibfnamefont {T.}~\bibnamefont {Ishida}}, \bibinfo {author} {\bibfnamefont {T.}~\bibnamefont {Nogami}}, \bibinfo {author} {\bibfnamefont {J.}~\bibnamefont {Kreitlow}}, \bibinfo {author} {\bibfnamefont {D.}~\bibnamefont {Baabe}}, \bibinfo {author} {\bibfnamefont {F.~J.}\ \bibnamefont {Litterst}}, \bibinfo {author} {\bibfnamefont {S.}~\bibnamefont {S\"ullow}}, \bibinfo {author} {\bibfnamefont {H.-H.}\ \bibnamefont {Klauss}}, \ and\ \bibinfo {author} {\bibfnamefont {K.}~\bibnamefont {Doll}},\ }\href {\doibase 10.1103/PhysRevB.69.134427} {\bibfield  {journal} {\bibinfo  {journal} {Phys. Rev. B}\ }\textbf {\bibinfo {volume} {69}},\ \bibinfo {pages} {134427} (\bibinfo {year} {2004})}\BibitemShut {NoStop}%
\bibitem [{\citenamefont {Kreitlow}\ \emph {et~al.}(2005)\citenamefont {Kreitlow}, \citenamefont {Menzel}, \citenamefont {Wolter}, \citenamefont {Schoenes}, \citenamefont {S\"ullow}, \citenamefont {Feyerherm},\ and\ \citenamefont {Doll}}]{XCl2_pressure}%
  \BibitemOpen
  \bibfield  {author} {\bibinfo {author} {\bibfnamefont {J.}~\bibnamefont {Kreitlow}}, \bibinfo {author} {\bibfnamefont {D.}~\bibnamefont {Menzel}}, \bibinfo {author} {\bibfnamefont {A.~U.~B.}\ \bibnamefont {Wolter}}, \bibinfo {author} {\bibfnamefont {J.}~\bibnamefont {Schoenes}}, \bibinfo {author} {\bibfnamefont {S.}~\bibnamefont {S\"ullow}}, \bibinfo {author} {\bibfnamefont {R.}~\bibnamefont {Feyerherm}}, \ and\ \bibinfo {author} {\bibfnamefont {K.}~\bibnamefont {Doll}},\ }\href {\doibase 10.1103/PhysRevB.72.134418} {\bibfield  {journal} {\bibinfo  {journal} {Phys. Rev. B}\ }\textbf {\bibinfo {volume} {72}},\ \bibinfo {pages} {134418} (\bibinfo {year} {2005})}\BibitemShut {NoStop}%
\bibitem [{\citenamefont {Pitcairn}\ \emph {et~al.}(2024)\citenamefont {Pitcairn}, \citenamefont {Ongkiko}, \citenamefont {Iliceto}, \citenamefont {Speakman}, \citenamefont {Calder}, \citenamefont {Cochran}, \citenamefont {Paddison}, \citenamefont {Liu}, \citenamefont {Argent}, \citenamefont {Morris} \emph {et~al.}}]{XCl2L_Jem_2024}%
  \BibitemOpen
  \bibfield  {author} {\bibinfo {author} {\bibfnamefont {J.}~\bibnamefont {Pitcairn}}, \bibinfo {author} {\bibfnamefont {M.~A.}\ \bibnamefont {Ongkiko}}, \bibinfo {author} {\bibfnamefont {A.}~\bibnamefont {Iliceto}}, \bibinfo {author} {\bibfnamefont {P.}~\bibnamefont {Speakman}}, \bibinfo {author} {\bibfnamefont {S.}~\bibnamefont {Calder}}, \bibinfo {author} {\bibfnamefont {M.}~\bibnamefont {Cochran}}, \bibinfo {author} {\bibfnamefont {J.}~\bibnamefont {Paddison}}, \bibinfo {author} {\bibfnamefont {C.}~\bibnamefont {Liu}}, \bibinfo {author} {\bibfnamefont {S.}~\bibnamefont {Argent}}, \bibinfo {author} {\bibfnamefont {A.}~\bibnamefont {Morris}},  \emph {et~al.},\ }\href {\doibase 10.26434/chemrxiv-2024-nvj38} {\bibfield  {journal} {\bibinfo  {journal} {ChemRxiv}\ } (\bibinfo {year} {2024}),\ 10.26434/chemrxiv-2024-nvj38}\BibitemShut {NoStop}%
\bibitem [{sup()}]{supplementary}%
  \BibitemOpen
  \href@noop {} {}\bibinfo {note} {See Supplemental Material at [URL will be inserted by publisher].}\BibitemShut {Stop}%
\bibitem [{\citenamefont {Fisher}(1962)}]{fisher_1962}%
  \BibitemOpen
  \bibfield  {author} {\bibinfo {author} {\bibfnamefont {M.~E.}\ \bibnamefont {Fisher}},\ }\href {\doibase 10.1080/14786436208213705} {\bibfield  {journal} {\bibinfo  {journal} {Philosophical Magazine}\ }\textbf {\bibinfo {volume} {7}},\ \bibinfo {pages} {1731} (\bibinfo {year} {1962})}\BibitemShut {NoStop}%
\bibitem [{\citenamefont {Chapon}\ \emph {et~al.}(2011)\citenamefont {Chapon}, \citenamefont {Manuel}, \citenamefont {Radaelli}, \citenamefont {Benson}, \citenamefont {Perrott}, \citenamefont {Ansell}, \citenamefont {Rhodes}, \citenamefont {Raspino}, \citenamefont {Duxbury}, \citenamefont {Spill} \emph {et~al.}}]{Wish}%
  \BibitemOpen
  \bibfield  {author} {\bibinfo {author} {\bibfnamefont {L.~C.}\ \bibnamefont {Chapon}}, \bibinfo {author} {\bibfnamefont {P.}~\bibnamefont {Manuel}}, \bibinfo {author} {\bibfnamefont {P.~G.}\ \bibnamefont {Radaelli}}, \bibinfo {author} {\bibfnamefont {C.}~\bibnamefont {Benson}}, \bibinfo {author} {\bibfnamefont {L.}~\bibnamefont {Perrott}}, \bibinfo {author} {\bibfnamefont {S.}~\bibnamefont {Ansell}}, \bibinfo {author} {\bibfnamefont {N.~J.}\ \bibnamefont {Rhodes}}, \bibinfo {author} {\bibfnamefont {D.}~\bibnamefont {Raspino}}, \bibinfo {author} {\bibfnamefont {D.}~\bibnamefont {Duxbury}}, \bibinfo {author} {\bibfnamefont {E.}~\bibnamefont {Spill}},  \emph {et~al.},\ }\href {https://doi.org/10.1080/10448632.2011.569650} {\bibfield  {journal} {\bibinfo  {journal} {Neutron News}\ }\textbf {\bibinfo {volume} {22}},\ \bibinfo {pages} {22} (\bibinfo {year} {2011})}\BibitemShut {NoStop}%
\bibitem [{\citenamefont {Rodr{\'\i}guez-Carvajal}(1993)}]{Fullprof}%
  \BibitemOpen
  \bibfield  {author} {\bibinfo {author} {\bibfnamefont {J.}~\bibnamefont {Rodr{\'\i}guez-Carvajal}},\ }\href {https://doi.org/10.1016/0921-4526(93)90108-I} {\bibfield  {journal} {\bibinfo  {journal} {Physica B: Condensed Matter}\ }\textbf {\bibinfo {volume} {192}},\ \bibinfo {pages} {55} (\bibinfo {year} {1993})}\BibitemShut {NoStop}%
\bibitem [{Iso()}]{Isodistort}%
  \BibitemOpen
  \href@noop {} {}\bibinfo {note} {H. T. Stokes, D. M. Hatch, and B. J. Campbell, ISODISTORT, ISOTROPY Software Suite, \url{iso.byu.edu},}\BibitemShut {NoStop}%
\bibitem [{\citenamefont {Campbell}\ \emph {et~al.}(2006)\citenamefont {Campbell}, \citenamefont {Stokes}, \citenamefont {Tanner},\ and\ \citenamefont {Hatch}}]{Isodistort_2006}%
  \BibitemOpen
  \bibfield  {author} {\bibinfo {author} {\bibfnamefont {B.~J.}\ \bibnamefont {Campbell}}, \bibinfo {author} {\bibfnamefont {H.~T.}\ \bibnamefont {Stokes}}, \bibinfo {author} {\bibfnamefont {D.~E.}\ \bibnamefont {Tanner}}, \ and\ \bibinfo {author} {\bibfnamefont {D.~M.}\ \bibnamefont {Hatch}},\ }\href {https://doi.org/10.1107/S0021889806014075} {\bibfield  {journal} {\bibinfo  {journal} {Journal of Applied Crystallography}\ }\textbf {\bibinfo {volume} {39}},\ \bibinfo {pages} {607} (\bibinfo {year} {2006})}\BibitemShut {NoStop}%
\bibitem [{Note1()}]{Note1}%
  \BibitemOpen
  \bibinfo {note} {The hard-axis of one site aligns within the easy-plane of the second site because of the exact $90^{\circ }$ rotation of the Ni(II) between neighbouring sites. For rotation by an angle which is not a multiple of $90^{\circ }$, saturation phase transitions will not occur for any directions perpendicular to the pseudo-easy-axis.}\BibitemShut {Stop}%
\end{thebibliography}%


\begin{thebibliography}{12}%
\makeatletter
\providecommand \@ifxundefined [1]{%
 \@ifx{#1\undefined}
}%
\providecommand \@ifnum [1]{%
 \ifnum #1\expandafter \@firstoftwo
 \else \expandafter \@secondoftwo
 \fi
}%
\providecommand \@ifx [1]{%
 \ifx #1\expandafter \@firstoftwo
 \else \expandafter \@secondoftwo
 \fi
}%
\providecommand \natexlab [1]{#1}%
\providecommand \enquote  [1]{``#1''}%
\providecommand \bibnamefont  [1]{#1}%
\providecommand \bibfnamefont [1]{#1}%
\providecommand \citenamefont [1]{#1}%
\providecommand \href@noop [0]{\@secondoftwo}%
\providecommand \href [0]{\begingroup \@sanitize@url \@href}%
\providecommand \@href[1]{\@@startlink{#1}\@@href}%
\providecommand \@@href[1]{\endgroup#1\@@endlink}%
\providecommand \@sanitize@url [0]{\catcode `\\12\catcode `\$12\catcode `\&12\catcode `\#12\catcode `\^12\catcode `\_12\catcode `\%12\relax}%
\providecommand \@@startlink[1]{}%
\providecommand \@@endlink[0]{}%
\providecommand \url  [0]{\begingroup\@sanitize@url \@url }%
\providecommand \@url [1]{\endgroup\@href {#1}{\urlprefix }}%
\providecommand \urlprefix  [0]{URL }%
\providecommand \Eprint [0]{\href }%
\providecommand \doibase [0]{http://dx.doi.org/}%
\providecommand \selectlanguage [0]{\@gobble}%
\providecommand \bibinfo  [0]{\@secondoftwo}%
\providecommand \bibfield  [0]{\@secondoftwo}%
\providecommand \translation [1]{[#1]}%
\providecommand \BibitemOpen [0]{}%
\providecommand \bibitemStop [0]{}%
\providecommand \bibitemNoStop [0]{.\EOS\space}%
\providecommand \EOS [0]{\spacefactor3000\relax}%
\providecommand \BibitemShut  [1]{\csname bibitem#1\endcsname}%
\let\auto@bib@innerbib\@empty
\bibitem [{Note1()}]{Note1}%
  \BibitemOpen
  \bibinfo {note} {NiCl$_{2}$(pym) exhibits a large magnetic hysteresis~\cite {XCl2L_Jem_2024}.}\BibitemShut {Stop}%
\bibitem [{Cry()}]{CrysAlisPRO}%
  \BibitemOpen
  \href@noop {} {\enquote {\bibinfo {title} {Crysalispro},}\ }\bibinfo {howpublished} {CrysAlisPRO, Oxford Diffraction /Agilent Technologies UK Ltd, Yarnton, England}\BibitemShut {NoStop}%
\bibitem [{\citenamefont {Sheldrick}(2015)}]{Sheldrick}%
  \BibitemOpen
  \bibfield  {author} {\bibinfo {author} {\bibfnamefont {G.~M.}\ \bibnamefont {Sheldrick}},\ }\href {\doibase 10.1107/S2053229614024218} {\bibfield  {journal} {\bibinfo  {journal} {Acta Crystallographica Section C}\ }\textbf {\bibinfo {volume} {71}},\ \bibinfo {pages} {3} (\bibinfo {year} {2015})}\BibitemShut {NoStop}%
\bibitem [{\citenamefont {Bourhis}\ \emph {et~al.}(2015)\citenamefont {Bourhis}, \citenamefont {Dolomanov}, \citenamefont {Gildea}, \citenamefont {Howard},\ and\ \citenamefont {Puschmann}}]{Bourhis_2015}%
  \BibitemOpen
  \bibfield  {author} {\bibinfo {author} {\bibfnamefont {L.~J.}\ \bibnamefont {Bourhis}}, \bibinfo {author} {\bibfnamefont {O.~V.}\ \bibnamefont {Dolomanov}}, \bibinfo {author} {\bibfnamefont {R.~J.}\ \bibnamefont {Gildea}}, \bibinfo {author} {\bibfnamefont {J.~A.}\ \bibnamefont {Howard}}, \ and\ \bibinfo {author} {\bibfnamefont {H.}~\bibnamefont {Puschmann}},\ }\href {https://journals.iucr.org/a/issues/2015/01/00/pc5043/index.html} {\bibfield  {journal} {\bibinfo  {journal} {Acta Crystallographica Section A: Foundations and Advances}\ }\textbf {\bibinfo {volume} {71}},\ \bibinfo {pages} {59} (\bibinfo {year} {2015})}\BibitemShut {NoStop}%
\bibitem [{\citenamefont {Dolomanov}\ \emph {et~al.}(2009)\citenamefont {Dolomanov}, \citenamefont {Bourhis}, \citenamefont {Gildea}, \citenamefont {Howard},\ and\ \citenamefont {Puschmann}}]{Dolomanov_2009}%
  \BibitemOpen
  \bibfield  {author} {\bibinfo {author} {\bibfnamefont {O.~V.}\ \bibnamefont {Dolomanov}}, \bibinfo {author} {\bibfnamefont {L.~J.}\ \bibnamefont {Bourhis}}, \bibinfo {author} {\bibfnamefont {R.~J.}\ \bibnamefont {Gildea}}, \bibinfo {author} {\bibfnamefont {J.~A.}\ \bibnamefont {Howard}}, \ and\ \bibinfo {author} {\bibfnamefont {H.}~\bibnamefont {Puschmann}},\ }\href {https://doi.org/10.1107/S0021889808042726} {\bibfield  {journal} {\bibinfo  {journal} {Journal of applied crystallography}\ }\textbf {\bibinfo {volume} {42}},\ \bibinfo {pages} {339} (\bibinfo {year} {2009})}\BibitemShut {NoStop}%
\bibitem [{\citenamefont {Saha}\ \emph {et~al.}(2022)\citenamefont {Saha}, \citenamefont {Nia},\ and\ \citenamefont {Rodr{\'\i}guez}}]{Saha_2022}%
  \BibitemOpen
  \bibfield  {author} {\bibinfo {author} {\bibfnamefont {A.}~\bibnamefont {Saha}}, \bibinfo {author} {\bibfnamefont {S.~S.}\ \bibnamefont {Nia}}, \ and\ \bibinfo {author} {\bibfnamefont {J.~A.}\ \bibnamefont {Rodr{\'\i}guez}},\ }\href {https://doi.org/10.1021/acs.chemrev.1c00879} {\bibfield  {journal} {\bibinfo  {journal} {Chemical Reviews}\ }\textbf {\bibinfo {volume} {122}},\ \bibinfo {pages} {13883} (\bibinfo {year} {2022})}\BibitemShut {NoStop}%
\bibitem [{\citenamefont {Klar}\ \emph {et~al.}(2023)\citenamefont {Klar}, \citenamefont {Krysiak}, \citenamefont {Xu}, \citenamefont {Steciuk}, \citenamefont {Cho}, \citenamefont {Zou},\ and\ \citenamefont {Palatinus}}]{Klar_2023}%
  \BibitemOpen
  \bibfield  {author} {\bibinfo {author} {\bibfnamefont {P.~B.}\ \bibnamefont {Klar}}, \bibinfo {author} {\bibfnamefont {Y.}~\bibnamefont {Krysiak}}, \bibinfo {author} {\bibfnamefont {H.}~\bibnamefont {Xu}}, \bibinfo {author} {\bibfnamefont {G.}~\bibnamefont {Steciuk}}, \bibinfo {author} {\bibfnamefont {J.}~\bibnamefont {Cho}}, \bibinfo {author} {\bibfnamefont {X.}~\bibnamefont {Zou}}, \ and\ \bibinfo {author} {\bibfnamefont {L.}~\bibnamefont {Palatinus}},\ }\href {https://www.nature.com/articles/s41557-023-01186-1} {\bibfield  {journal} {\bibinfo  {journal} {Nature Chemistry}\ }\textbf {\bibinfo {volume} {15}},\ \bibinfo {pages} {848} (\bibinfo {year} {2023})}\BibitemShut {NoStop}%
\bibitem [{\citenamefont {Liu}\ \emph {et~al.}(2016)\citenamefont {Liu}, \citenamefont {Goddard}, \citenamefont {Singleton}, \citenamefont {Brambleby}, \citenamefont {Foronda}, \citenamefont {M\"{o}ller}, \citenamefont {Kohama}, \citenamefont {Ghannadzadeh}, \citenamefont {Ardavan}, \citenamefont {Blundell} \emph {et~al.}}]{Liu_2016_correlations}%
  \BibitemOpen
  \bibfield  {author} {\bibinfo {author} {\bibfnamefont {J.}~\bibnamefont {Liu}}, \bibinfo {author} {\bibfnamefont {P.~A.}\ \bibnamefont {Goddard}}, \bibinfo {author} {\bibfnamefont {J.}~\bibnamefont {Singleton}}, \bibinfo {author} {\bibfnamefont {J.}~\bibnamefont {Brambleby}}, \bibinfo {author} {\bibfnamefont {F.}~\bibnamefont {Foronda}}, \bibinfo {author} {\bibfnamefont {J.~S.}\ \bibnamefont {M\"{o}ller}}, \bibinfo {author} {\bibfnamefont {Y.}~\bibnamefont {Kohama}}, \bibinfo {author} {\bibfnamefont {S.}~\bibnamefont {Ghannadzadeh}}, \bibinfo {author} {\bibfnamefont {A.}~\bibnamefont {Ardavan}}, \bibinfo {author} {\bibfnamefont {S.~J.}\ \bibnamefont {Blundell}},  \emph {et~al.},\ }\href {https://doi.org/10.1021/acs.inorgchem.5b02991} {\bibfield  {journal} {\bibinfo  {journal} {Inorganic Chemistry}\ }\textbf {\bibinfo {volume} {55}},\ \bibinfo {pages} {3515} (\bibinfo {year} {2016})}\BibitemShut {NoStop}%
\bibitem [{\citenamefont {Blackmore}\ \emph {et~al.}(2022)\citenamefont {Blackmore}, \citenamefont {Curley}, \citenamefont {Williams}, \citenamefont {Vaidya}, \citenamefont {Singleton}, \citenamefont {Birnbaum}, \citenamefont {Ozarowski}, \citenamefont {Schlueter}, \citenamefont {Chen}, \citenamefont {Gillon}, \citenamefont {Goukassov}, \citenamefont {Kibalin}, \citenamefont {Villa}, \citenamefont {Villa}, \citenamefont {Manson},\ and\ \citenamefont {Goddard}}]{Blackmore-correlations}%
  \BibitemOpen
  \bibfield  {author} {\bibinfo {author} {\bibfnamefont {W.~J.~A.}\ \bibnamefont {Blackmore}}, \bibinfo {author} {\bibfnamefont {S.~P.~M.}\ \bibnamefont {Curley}}, \bibinfo {author} {\bibfnamefont {R.~C.}\ \bibnamefont {Williams}}, \bibinfo {author} {\bibfnamefont {S.}~\bibnamefont {Vaidya}}, \bibinfo {author} {\bibfnamefont {J.}~\bibnamefont {Singleton}}, \bibinfo {author} {\bibfnamefont {S.}~\bibnamefont {Birnbaum}}, \bibinfo {author} {\bibfnamefont {A.}~\bibnamefont {Ozarowski}}, \bibinfo {author} {\bibfnamefont {J.~A.}\ \bibnamefont {Schlueter}}, \bibinfo {author} {\bibfnamefont {Y.-S.}\ \bibnamefont {Chen}}, \bibinfo {author} {\bibfnamefont {B.}~\bibnamefont {Gillon}}, \bibinfo {author} {\bibfnamefont {A.}~\bibnamefont {Goukassov}}, \bibinfo {author} {\bibfnamefont {I.}~\bibnamefont {Kibalin}}, \bibinfo {author} {\bibfnamefont {D.~Y.}\ \bibnamefont {Villa}}, \bibinfo {author} {\bibfnamefont {J.~A.}\ \bibnamefont {Villa}}, \bibinfo {author} {\bibfnamefont {J.~L.}\ \bibnamefont {Manson}}, \ and\ \bibinfo
  {author} {\bibfnamefont {P.~A.}\ \bibnamefont {Goddard}},\ }\href {\doibase 10.1021/acs.inorgchem.1c02483} {\bibfield  {journal} {\bibinfo  {journal} {Inorganic Chemistry}\ }\textbf {\bibinfo {volume} {61}},\ \bibinfo {pages} {141} (\bibinfo {year} {2022})}\BibitemShut {NoStop}%
\bibitem [{\citenamefont {Chapon}\ \emph {et~al.}(2011)\citenamefont {Chapon}, \citenamefont {Manuel}, \citenamefont {Radaelli}, \citenamefont {Benson}, \citenamefont {Perrott}, \citenamefont {Ansell}, \citenamefont {Rhodes}, \citenamefont {Raspino}, \citenamefont {Duxbury}, \citenamefont {Spill} \emph {et~al.}}]{Wish}%
  \BibitemOpen
  \bibfield  {author} {\bibinfo {author} {\bibfnamefont {L.~C.}\ \bibnamefont {Chapon}}, \bibinfo {author} {\bibfnamefont {P.}~\bibnamefont {Manuel}}, \bibinfo {author} {\bibfnamefont {P.~G.}\ \bibnamefont {Radaelli}}, \bibinfo {author} {\bibfnamefont {C.}~\bibnamefont {Benson}}, \bibinfo {author} {\bibfnamefont {L.}~\bibnamefont {Perrott}}, \bibinfo {author} {\bibfnamefont {S.}~\bibnamefont {Ansell}}, \bibinfo {author} {\bibfnamefont {N.~J.}\ \bibnamefont {Rhodes}}, \bibinfo {author} {\bibfnamefont {D.}~\bibnamefont {Raspino}}, \bibinfo {author} {\bibfnamefont {D.}~\bibnamefont {Duxbury}}, \bibinfo {author} {\bibfnamefont {E.}~\bibnamefont {Spill}},  \emph {et~al.},\ }\href {https://doi.org/10.1080/10448632.2011.569650} {\bibfield  {journal} {\bibinfo  {journal} {Neutron News}\ }\textbf {\bibinfo {volume} {22}},\ \bibinfo {pages} {22} (\bibinfo {year} {2011})}\BibitemShut {NoStop}%
\bibitem [{\citenamefont {Rodr{\'\i}guez-Carvajal}(1993)}]{Fullprof}%
  \BibitemOpen
  \bibfield  {author} {\bibinfo {author} {\bibfnamefont {J.}~\bibnamefont {Rodr{\'\i}guez-Carvajal}},\ }\href {https://doi.org/10.1016/0921-4526(93)90108-I} {\bibfield  {journal} {\bibinfo  {journal} {Physica B: Condensed Matter}\ }\textbf {\bibinfo {volume} {192}},\ \bibinfo {pages} {55} (\bibinfo {year} {1993})}\BibitemShut {NoStop}%
\bibitem [{\citenamefont {Pitcairn}\ \emph {et~al.}(2024)\citenamefont {Pitcairn}, \citenamefont {Ongkiko}, \citenamefont {Iliceto}, \citenamefont {Speakman}, \citenamefont {Calder}, \citenamefont {Cochran}, \citenamefont {Paddison}, \citenamefont {Liu}, \citenamefont {Argent}, \citenamefont {Morris} \emph {et~al.}}]{XCl2L_Jem_2024}%
  \BibitemOpen
  \bibfield  {author} {\bibinfo {author} {\bibfnamefont {J.}~\bibnamefont {Pitcairn}}, \bibinfo {author} {\bibfnamefont {M.~A.}\ \bibnamefont {Ongkiko}}, \bibinfo {author} {\bibfnamefont {A.}~\bibnamefont {Iliceto}}, \bibinfo {author} {\bibfnamefont {P.}~\bibnamefont {Speakman}}, \bibinfo {author} {\bibfnamefont {S.}~\bibnamefont {Calder}}, \bibinfo {author} {\bibfnamefont {M.}~\bibnamefont {Cochran}}, \bibinfo {author} {\bibfnamefont {J.}~\bibnamefont {Paddison}}, \bibinfo {author} {\bibfnamefont {C.}~\bibnamefont {Liu}}, \bibinfo {author} {\bibfnamefont {S.}~\bibnamefont {Argent}}, \bibinfo {author} {\bibfnamefont {A.}~\bibnamefont {Morris}},  \emph {et~al.},\ }\href {\doibase 10.26434/chemrxiv-2024-nvj38} {\bibfield  {journal} {\bibinfo  {journal} {ChemRxiv}\ } (\bibinfo {year} {2024}),\ 10.26434/chemrxiv-2024-nvj38}\BibitemShut {NoStop}%
\end{thebibliography}%

\end{document}


\title{Supplementary Material accompanying \\ Pseudo-easy-axis anisotropy in antiferromagnetic $S=1$ diamond lattice systems}

\author{S. Vaidya}
\email{s.vaidya@warwick.ac.uk}
\affiliation{Department of Physics, University of Warwick, Gibbet Hill Road, Coventry, CV4 7AL, UK}
\author{A. Hern\'{a}ndez-Meli\'{a}n}
\affiliation{Department of Physics, Durham University, South Road, Durham, DH1 3LE, United Kingdom}
\author{J. P. Tidey}
\affiliation{Department of Physics, University of Warwick, Gibbet Hill Road, Coventry, CV4 7AL, UK}
\author{S. P. M. Curley}
\author{S. Sharma}
\affiliation{Department of Physics, University of Warwick, Gibbet Hill Road, Coventry, CV4 7AL, UK}
\author{P. Manuel}
\affiliation{ISIS Pulsed Neutron Source, STFC Rutherford Appleton Laboratory, Didcot, Oxfordshire OX11 0QX, United Kingdom}
\author{C. Wang}
\affiliation{Laboratory for Muon Spin Spectroscopy, Paul Scherrer Institute, 5232 Villigen, Switzerland}
\author{G. L. Hannaford}
\affiliation{Department of Physics, Durham University, South Road, Durham, DH1 3LE, United Kingdom}
\author{S. J. Blundell}
\affiliation{Department of Physics, Clarendon Laboratory, University of Oxford, Parks Road, Oxford, OX1 3PU, United Kingdom}
\author{Z. E. Manson}
\author{J. L. Manson}\thanks{Deceased 7 June 2023.}
\affiliation{Department of Chemistry and Biochemistry, Eastern Washington University, Cheney, Washington 99004, USA}
\author{J. Singleton}
\affiliation{National High Magnetic Field Laboratory (NHMFL), Los Alamos National Laboratory, Los Alamos, New Mexico 87545, USA}
\author{T. Lancaster}
\affiliation{Department of Physics, Durham University, South Road, Durham, DH1 3LE, United Kingdom}
\author{R. D. Johnson}
\affiliation{Department of Physics and Astronomy, University College London, London, UK}
\affiliation{London Centre for Nanotechnology and Department of Physics and Astronomy, University College London, London WC1E 6BT, UK}
\author{P. A. Goddard}
\email{p.goddard@warwick.ac.uk}
\affiliation{Department of Physics, University of Warwick, Gibbet Hill Road, Coventry, CV4 7AL, UK}

\maketitle
\tableofcontents
\section{Further Experimental Details}
\subsection{Synthesis}
All chemical reagents were obtained from commercial sources and used as received. In a typical synthesis of NiCl2(pyrimidine)2, 1.500 g of NiCl2 was dissolved in 2 ml of H20 and, separately, 1.0110 g of pyrimidine (1.0110 g) was dissolved in 2 ml of H2O. The NiCl2 solution was slowly mixed with the ligand solution in 5 ml of heated ethanol to afford a solution. When allowed to slowly evaporate a fine powder was recovered. For a typical synthesis of NiBr2(pyrimidine)2, 0.6000 g of NiBr2 was dissolved in 3 ml of H20 and 3 ml ethanol. This was slowly mixed with 0.4392 g of pyrimidine dissolved in 2 ml of H2O and 3 ml ethanol and an ocean blue solution was obtained. Evaporation yielded a fine powder.
\subsection{Magnetometry}
\paragraph{SQUID}
\begin{figure}
    \centering
    \includegraphics[width=\linewidth]{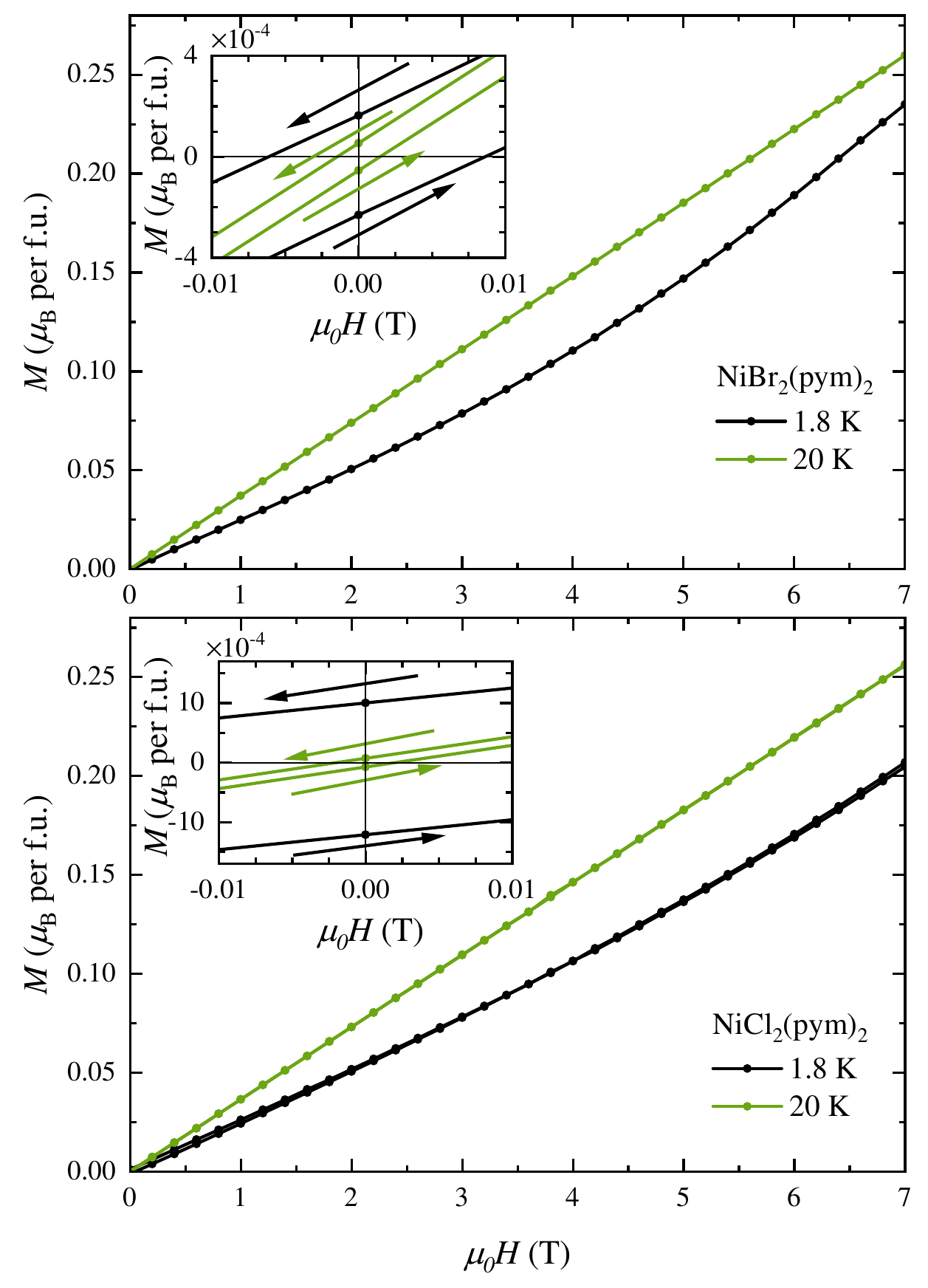}
    \caption{Powder magnetization measurements of Ni$X_{2}$(pym)$_{2}$ for (a) $X$\,=Cl and (b) $X$\,=Br performed using SQUID magnetometer. The insets are zoomed in around $\mu_{0}H = 0$\,T and show the negligible hysteresis.}
    \label{fig: mag_hys}
\end{figure}
Measurements of $\chi(T)$ are performed using a Quantum Design MPMS XL SQUID magnetometer. Powder samples of Ni$X_{2}$(pym)$_{2}$ are suspended in Vaseline to mitigate the movements of grains in the field and the Vaseline-sample mix is held within a gelatine capsule with a low magnetic background. The gelatine capsule is then placed in a plastic drinking straw for the measurement. Samples are cooled to $T=1.8$\,K in zero-field and zero-field-cooled (ZFC) measurements are performed on warming to $T=300$\,K with a constant applied field of $\mu_{0}H = 0.1$\,T. Field-cooled measurements are then performed on cooling back down to $T=1.8$\,K. $M(H)$ measurements at constant temperatures of $T=1.8$\,K and $T = 20$\,K are performed for the ZFC samples. For these measurements a field-sweeps sequence of $0\rightarrow7\rightarrow-7\rightarrow7\rightarrow0$\,T is used to look for remanent magnetisation at $\mu_{0}H = 0$\,T. However, as shown in Fig.~\ref{fig: mag_hys} only a very small remanent $M(0)$ of $0.001$\,$\mu_{\text{B}}$ and $0.00016$\,$\mu_{\text{B}}$\,per Ni(II) ion in $X$\,=Cl and Br respectively, were observed at $T=1.8$\,K.

\paragraph{Pulsed-field} Isothermal pulsed-field magnetization measurements are performed at the National High Magnetic Field Laboratory in Los Alamos, USA. Fields of up to $65$\,T with a typical rise time of $\approx10$\,ms are used. Polycrystalline samples are packed into a PCTFE ampoule (inner diameter $1.0$\,mm) and sealed with vacuum grease to prevent sample movement. The ampoule can be moved in and out of a $1500$-turn, $1.5$\,mm bore, $1.5$\,mm-long compensated-coil susceptometer constructed from $50$-gauge high-purity copper wire. When the sample is in the coil, the voltage induced in the coil is proportional to the rate of change in $M$ over time (d$M$/d$t$). The signal is integrated and the background data, measured with an empty coil under the same conditions, is subtracted to obtain $M(H)$. The magnetic field value is measured using a coaxial $10$-turn coil and calibrated using observations of de Haas-van Alphen oscillations arising from the copper coils of the susceptometer. A $^{3}$He cryostat provides temperature control and is used to attain temperatures down to $500$\,mK. The $M(H)$ measured in pulsed-fields are normalised using $T=1.8$\,K data measured in the SQUID magnetometer.

\subsection{Crystal structure determination}

\paragraph{Powder X-ray diffraction} 
\begin{figure}
    \centering
    \includegraphics[width= \linewidth]{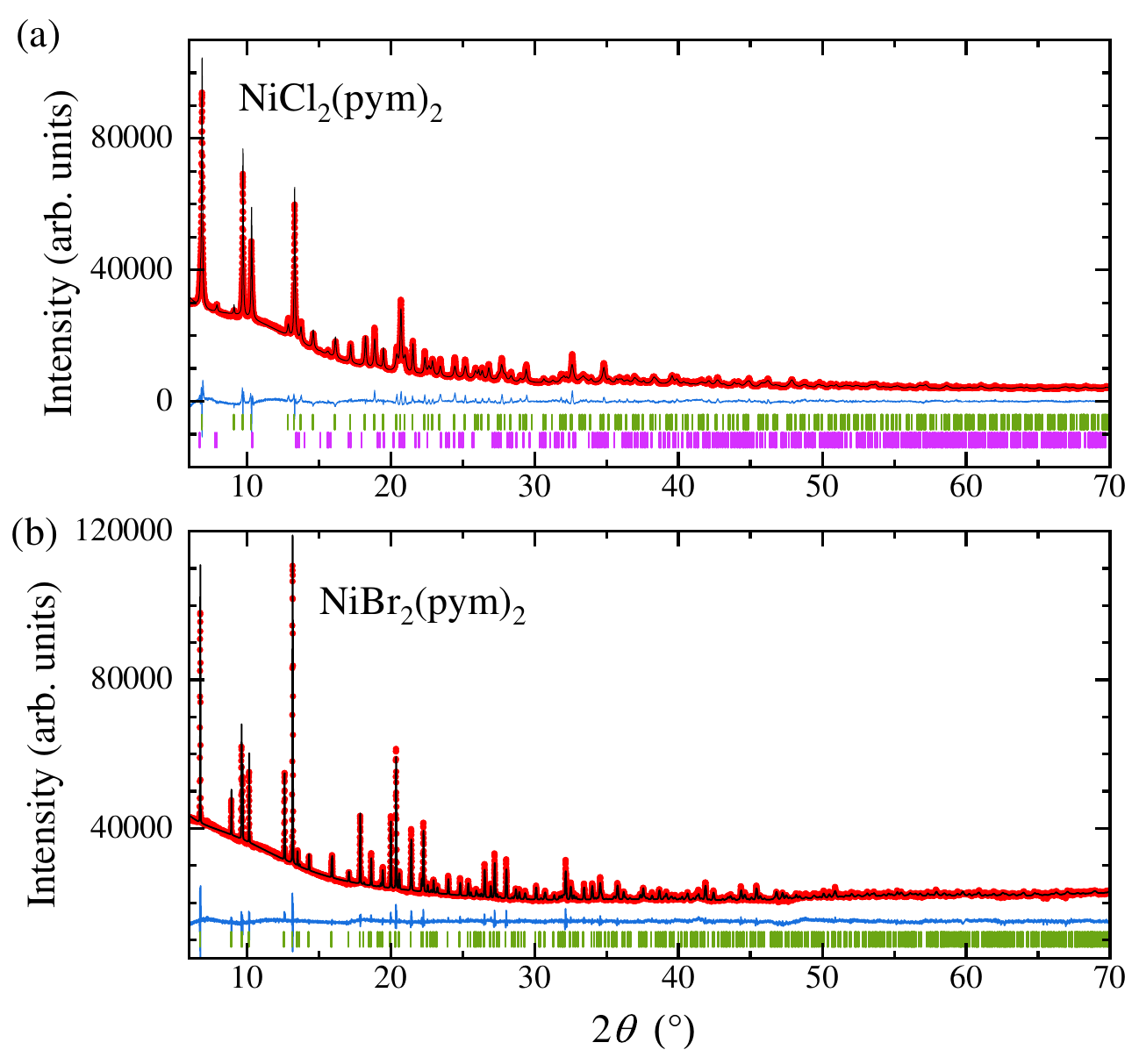}
    \caption{Refinement of the powder synchrotron X-ray diffraction measurements, at $T=120$\,K, for Ni$X_{2}$(pym)$_{2}$, where (a)$X\,=Cl$ and (b)$X\,=Br$. In both the data are shown as red circles, the fitted model as black lines, the difference between the model and data as blue lines and green tick marks indicate the Bragg positions of the $I4_{1}22$ model. The purple ticks in (a) mark the Bragg position of NiCl$_{2}$(pym) impurity phase with orthorhombic space group $Pmma$. The intensity of the impurity peaks imply it is $<10\%$ of the sample and has negligible effect on the measured magnetic properties of NiCl$_{2}$(pym)$_{2}$.}
    \label{fig: NiX2_PXRD}
\end{figure}

PXRD measurements are performed using synchrotron radiation $\lambda = 0.82398$\,$\si{\angstrom}$ at the I11 beamline of the Diamond Light Source, Didcot, UK. Samples are ground into fine powder and used to fill quartz capillary tubes with an inner diameter of $0.5$\,mm. Data is collected using a MYTHEN detector at $T = 120$\,K. Rietveld analysis of the data is performed in the TOPAS software. Table~\ref{tab: powder_crystal} lists the structural and refinement parameters. For refinement of the NiCl$_{2}$(pym)$_{2}$ data, a small amount of an orthorhombic $Pmma$, NiCl$_{2}$(pym) impurity phase was used to account for unindexed peaks. FeCl$_{2}$(pym)$_{2}$ is known to decompose to FeCl$_{2}$(pym) over the timescale of several years and as such presence of NiCl$_{2}$(pym) could be explicable by the slow decomposition of our nearly pure NiCl$_{2}$(pym)$_{2}$ sample. However, our bulk magnetometry\footnote{NiCl$_{2}$(pym) exhibits a large magnetic hysteresis~\cite{XCl2L_Jem_2024}.} and neutron scattering measurements, performed over a year before the PXRD and electron diffraction measurements, show no contributions from the impurity phase.

\begin{table*}
\caption{\label{tab: powder_crystal} Structural and refinement parameters for Ni$X_{2}$(pym)$_{2}$ ($X$\,=Cl, Br) determined through powder diffraction techniques.}
\begin{ruledtabular}
\begin{tabular}{cccc}
\textrm{Compound}& \multicolumn{2}{c}{NiCl$_{2}$(pym)$_{2}$}& \multicolumn{1}{c}{NiBr$_{2}$(pym)$_{2}$}\\
\colrule
Emp. formula                    & \multicolumn{2}{c}{NiCl$_{2}$C$_{8}$H$_{8}$N$_{2}$} & NiBr$_{2}$C$_{8}$H$_{8}$N$_{2}$ \\
Formula weight\,(g/mol)         &\multicolumn{2}{c}{$289.779$}  & $378.681$     \\
Crystal system, Space group     &\multicolumn{2}{c}{Tetragonal, $I4_{1}22$}&Tetragonal, $I4_{1}22$\\
Instrument                      & WISH          & I11           & I11           \\
Method                          & Powder Neutron & Powder X-ray & Powder X-ray  \\
$T$\,(K)                        & $20$          & $120$         & $120$         \\
$\lambda$\,$(\si{\angstrom})$   & --            & $0.82398$     & $0.82398$     \\
$a,b$\,$(\si{\angstrom})$       & $7.3426(2)$   & $7.3576(2)$   & $7.50690(6)$  \\
$c$\,$(\si{\angstrom})$         & $19.4841(9)$  & $19.4956(5)$  & $19.5378(2)$  \\
$V$\,$(\si{\angstrom}^{3})$     & $1050.45(7)$  & $1055.38(7)$  & $1101.03(2)$  \\
$Z$                             & $4$           & $4$           & $4$           \\
GOF or $\chi^{2}$\,$(\si{\angstrom}^{3})$ & $213.7$ & $4.1149$  & $2.4494$      \\
$R_{\text{exp}}$\,($\%$)        & $0.70$        & $1.5651$      & $1.0302$      \\
$R_{\text{WP}}$\,($\%$)         & $9.45$        & $4.2393$      & $1.5651$      \\
$R_{\text{Bragg}}$\,($\%$)      & $3.3738$      & $1.8590$      & $ 1.0695$     \\
\end{tabular}
\end{ruledtabular}
\end{table*}

\paragraph{Electron diffraction} 

For 3D-electron diffraction measurements, samples are lightly ground between two glass slides and dispersed onto copper-supported holey carbon TEM grid and loaded via a Gatan Elsa model 689 cryo holder into a Rigaku-Jeol XtaLAB Synergy-ED electron diffractometer, with a thermionic LaB$_{6}$ electron source operated at 200 kV and data recorded at a Rigaku HyPix-ED hybrid pixel array area detector.

In each case, data for a range of crystallites are surveyed at $120(2)$\,K using continuous rotation electron diffraction with a selected area of $\sim 2\,\mu$m diameter at the image plane using CrysAlisPRO (version 1.171.43.11a)~\cite{CrysAlisPRO}. The reported data are taken from representative datasets of optimum quality from crystallites of $\sim0.1-0.2$\,$\mu$m in dimensions collected over tilt ranges of $\sim~100^{\circ}$ with frame scan widths of $0.25^{\circ}$. Data in each case are indexed, integrated, and scaled using CrysAlisPRO (version 1.171.43.120a)~\cite{CrysAlisPRO} and SCALE3 ABSPACK implemented therein.

Structures are solved using ShelXT~\cite{Sheldrick} and refined using Olex2.refine~\cite{Bourhis_2015} NBeam approach to multiple diffraction as implemented in Olex2, version 1.5-ac6-018 (compiled 2024.02.14 svn.Rf8d729c for Rigaku Oxford Diffraction, GUI svn.r6920)~\cite{Dolomanov_2009} using published scattering factors~\cite{Saha_2022}. The correct hand of the refinement is determined using the Z-score approach~\cite{Klar_2023}.

C-H hydrogen atom positions are refined freely in the presence of a distance restraint to neutron values (1.08\,$\si{\angstrom}$) with riding isotropic displacement parameters, while a rigid bond restraint was applied over whole models to improve anisotropic displacement parameters and aid convergence. In the final stages of the refinement, unit cell parameters were imported to the models from the PXRD measurements, being taken as significantly more reliable than those obtained from ED as a consequence of the various distortions and instabilities inherent to the experiment. Further experimental details are provided in Table~\ref{tab: Elec_diff} with complete experimental, refinement, and structural information further contained in the deposited CIFs along with structure factors and embedded .RES files. The ED/PXRD derived structures are deposited and may be found at Deposition Numbers CCDC 2354990-2354994.

\begin{turnpage}
\begin{table*}
\caption{\label{tab: Elec_diff} Details of electron diffraction measurements for Ni$X_{2}$(pym)$_{2}$ ($X$\,=Cl, Br).}
\begin{ruledtabular}
\begin{tabular}{p{5cm}p{2.8cm}m{2.8cm}m{2.8cm}m{2.8cm}m{2.8cm}}
\textrm{Compound}& \multicolumn{3}{c}{NiCl$_{2}$(pym)$_{2}$}& \multicolumn{2}{c}{NiBr$_{2}$(pym)$_{2}$} \\
\colrule
Exp. No.                        &$1$       &$2$        &$3$        &$1$        &$2$        \\
\colrule
\multicolumn{6}{c}{Crystal data}\\
\colrule
$T$\,(K)                    &$120$ &$120$ &$120$ &$120$ &$120$ \\
Crystal system              &Tetragonal &Tetragonal &Tetragonal &Tetragonal &Tetragonal \\
Space group                 &$I4_{1}22$ &$I4_{1}22$ &$I4_{1}22$ &$I4_{1}22$ &$I4_{1}22$ \\
Cell (as used) $a, c\,(\si{\angstrom})$	&$7.3(3)$, $19.4(6)$&$7.3(2)$, $19.4(3)$&$7.29(16)$, $19.4(2)$&$7.45(10)$, $19.36(14)$&$7.45(13)$, $19.35(19)$\\
$V\,(\si{\angstrom}^{3})$ (as used)	&$1043(67)$	&$1031(44)$	&$1031(35)$	&$1075(22)$	&$1075(28)$\\
Cell (imported from PXRD) $a, c\,(\si{\angstrom})$	&$7.3576$, $19.4956 (5)$&$7.3576$, $19.4956 (5)$&$7.3576$, $19.4956 (5)$&$7.50690 (6)$, $19.5378 (2)$&$7.50690 (6)$, $19.5378 (2)$\\
$V\,(\si{\angstrom}^{3})$ (imported from PXRD)	&$1055.38(5)$&$1055.38(5)$&$1055.38(5)$&$1101.02(2)$&$1101.02(2)$\\
$Z$                         &4&4&4&4&4\\
\colrule
\multicolumn{6}{c}{Data collection}\\
\colrule
Tilt range (start, end) ($^{\circ}$)&$-37.75, 55.0$&$-60.0, 55.0$&$-60.0, 60.0$&$-67.5, 30.0$&$-40.0, 66.25$\\
Frame width ($^{\circ}$) &$0.25$    &$0.25$     &$0.25$     &$0.25$     &$0.25$\\
No. of measured, independent and observed [$I\geq2\sigma(I)$] reflections &$1809,1009,390$&	$2597,1456,589$&$4454,2060,871$&$2463,1315,800$&$3536,1738,877$\\
$R_{\text{int}}$                &$0.297$ &$0.288$  &$0.254$  &$0.163$ &$0.243$ \\
$(\sin(\theta/\lambda))_{\text{max}}\,(\si{\angstrom}^{-1})$&$0.591$&$0.619$&$0.751$&$0.633$&$0.709$\\
\colrule
\multicolumn{6}{c}{Refinement}\\
\colrule
\multirow{3}{*}{R indexes [$F^{2}\geq2\sigma(F^{2})$]}&R\,=\,$0.1101$ &R\,=\,$0.1260$ &R\,=\,$0.1115$   &R\,=\,$0.0961$ &R\,=\,$0.109$ \\
        &$w$R\,=\,$0.324$&$w$R\,=\,$0.392$&$w$R\,=\,$0.386$&$w$R\,=\,$0.267$&$w$R\,=\,$0.304$\\
        &$S$\,=\,$0.81$&$S$\,=\,$0.80$&$S$\,=\,$0.81$&$S$\,=\,$0.87$&$S$\,=\,$0.004$\\
Data/param./restr. &$461$/$42$/$45$ & $523$/$42$/$39$ & $882$/$42$/$33$ & $597$/$42$/$37$ & $782$/$42$/$33$\\
$\Delta\rho_{\text{max}}$, $\Delta\rho_{\text{max}}$\,e$\si{\angstrom}^{-3}$&$2.26$, $-2.15$&$3.89$, $-3.99$&$4.34$, $-3.64$&$2.53$, $-2.23$&$3.46$, $-2.94$\\
Z-score (noise adjusted, raw) &$11.69$, $5.32$&$14.61$, $6.92$&$18.17$, $8.20$&$13.55$, $7.64$&$17.06$, $8.59$\\
\end{tabular}
\end{ruledtabular}
\end{table*}
\end{turnpage}

\begin{table*}
\caption{\label{tab: bond_dist}%
Bond Lengths in Ni$X_{2}$(pym)$_{2}$ determined using electron diffraction measurements at $120$\,K. Ni---X, where $X$\,=\,Cl or Br represents the axial coordination bond lengths and Ni---N are the equatorial coordination bond lengths. [Ni---X]/[Ni---N] shows the ratio of axial and the equatorial coordination bond length. Secondary exchange interactions are along Ni---$X\cdot\cdot\cdot X$---Ni chains, for which $X\cdot\cdot\cdot X$ is halide-halide distances and Ni---$X\cdot\cdot\cdot X$---Ni is the separation of Ni ions along the weakly connected chain.}
\begin{ruledtabular}
\begin{tabular}{cccccc}
\textrm{}&
\textrm{Ni---X\,($\si{\angstrom}$)}&
\textrm{$X\cdot\cdot\cdot X$\,($\si{\angstrom}$)}&
\textrm{Ni---$X\cdot\cdot\cdot X$---Ni\,($\si{\angstrom}$)}&
\textrm{Ni---N\,($\si{\angstrom}$)}&
\textrm{[Ni---X]/[Ni---N]}\\\\
\colrule
NiCl$_{2}$(pym)$_{2}$ & 2.405(2) & 5.741(4) & 10.4052(3) & 2.140(4) & 1.124(5)\\
NiBr$_{2}$(pym)$_{2}$ & 2.567(1) & 5.480(3) & 10.6136(9) & 2.162(4) & 1.188(4)\\
\end{tabular}
\end{ruledtabular}
\end{table*}

Table~\ref{tab: bond_dist} provides the relevant bond distances, determined through electron diffraction measurements, for Ni$X_{2}$(pym)$_{2}$. In $X$\,=\,Cl, the equatorial bonds are $2.140(4)\,\si{\angstrom}$ long, the axial bonds are $2.405(2)\,\si{\angstrom}$ long and the axial to equatorial bond-length ratio is $1.124(5)$. The Ni(II) octahedra in $X$\,=\,Br exhibit a more pronounced elongation with equatorial bond-lengths of $2.162(4)\,\si{\angstrom}$ long, axial bonds are $2.567(1)\,\si{\angstrom}$ and a ratio is $1.188(4)$. The elongation is likely due to greater electronegativity of the Br ligands and is a trend previously explored in Ni$X_{2}$(pyz)$_{2}$~\cite{Liu_2016_correlations} and Ni$X_{2}$(3,5-lut)$_{4}$~\cite{Blackmore-correlations}.

\subsection{Muon-spin rotation}

Zero-field muon-spin rotation ($\mu^{+}$SR) measurements were carried out on polycrystalline samples of Ni$X_{2}$(pym)$_{2}$ ($X$\,=Cl, Br) using the GPS spectrometer at the Swiss Muon Source, Paul Scherrer Institue. For these measurements, samples are packed into Ag foil packets and mounted on a fork that is suspended in to the muon beam. 

In a $\mu^{+}$SR experiment spin-polarized positive muons are stopped in a target sample, where the muon usually occupies an interstitial position in the crystal. The observed property in the experiment is the time evolution of the muon spin polarization, the behaviour of which depends on the local magnetic field at the muon site. A quasistatic local field arising from magnetic order will typically cause a coherent precession of the muon spins. Each muon decays, with an average lifetime of $2.2$\,s, into two neutrinos and a positron, the latter particle being emitted preferentially along the instantaneous direction of the muon spin. Recording the time dependence of the positron emission directions therefore allows the determination of the spin-polarization of the ensemble of muons. In our experiments positrons are detected by detectors placed forward (F) and backward (B) of the initial muon polarization direction. Histograms $N_{\text{F}}(t)$ and $N_{\text{B}}(t)$ record the number of positrons detected in the two detectors as a function of time following the muon implantation. The quantity of interest is the decay positron asymmetry function, defined as
\begin{equation}
    A(t) = \frac{N_{\text{F}}(t) - \alpha_{\text{exp}}N_{\text{B}}(t)}{N_{\text{F}}(t) + \alpha_{\text{exp}}N_{\text{B}}(t)},
\end{equation}
where $\alpha_{\text{exp}}$ is an experimental calibration constant. $A(t)$ is proportional to the spin polarization of the muon ensemble.
\subsection{Elastic neutron scattering}

Elastic neutron scattering measurements are recorded on the WISH instrument at ISIS, Rutherford Appleton Laboratory, UK~\cite{Wish}. Powder samples of mass $0.98$\,g are loaded into a cylindrical vanadium can up to a height of $5$\,cm. An Oxford Instruments cryostat provides temperature control. Diffraction data is collected with long counting times of $7$\,hours at each temperature ($5$\,K and $20$\,K). Rietveld refinement of the nuclear structure is performed using the FULLPROF software~\cite{Fullprof} using the data collected across four detector banks, at $T=20$\,K. A description of the refinement and structural parameters and comparison to the PXRD parameters is given in Table~\ref{tab: powder_crystal}. Magnetic structure refinement was also performed using FULLPROF, on the magnetic diffraction obtained by subtracting the $20$\,K data from the $5$\,K data.

\section{Calculations}
\subsection{Mean-field calculations}
\begin{figure}
    \centering
    \includegraphics[width= \linewidth]{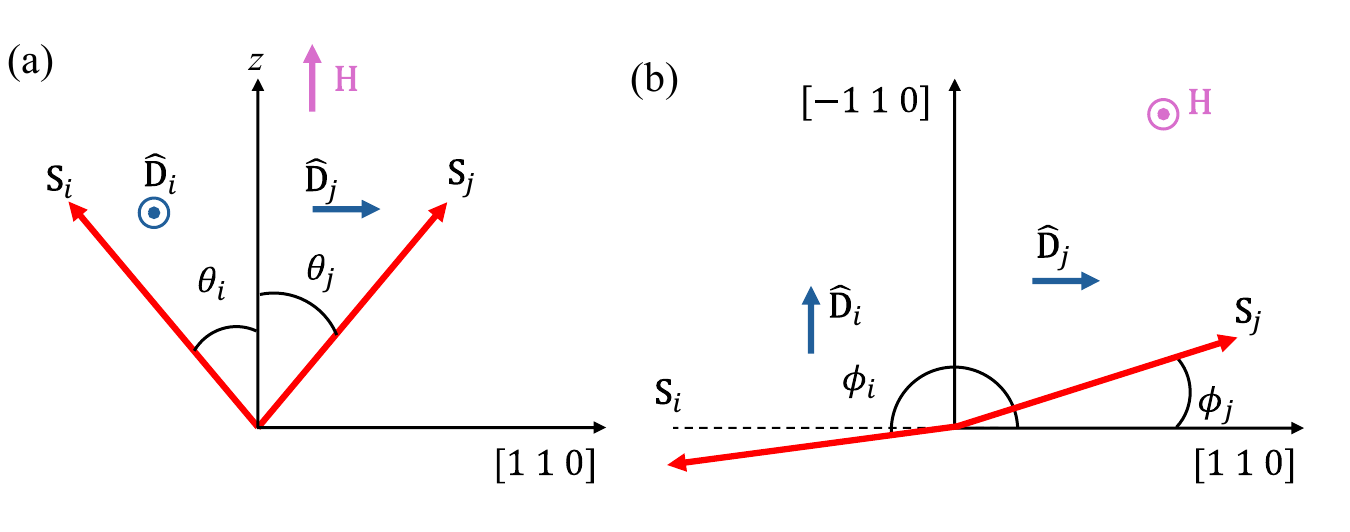}
    \caption{Classical configuration of the $i$ and $j$ sites in Ni$X_{2}$(3,5-lut)$_{4}$ within the spin-flop phase. The views down the [-1 1 0] and the [0 0 1] directions are depicted in (a) and (b) respectively. The relative directions of the single-ion anisotropy hard-axes, $\hat{\mathbf{D}}_{i,j}$ are shown as blue arrows and the purple arrows indicate the direction of the applied field $\mathbf{H}||$\,[0 0 1].}
    \label{fig: Flop_struct}
\end{figure}

The energy of the ground state determined using the Hamiltonian in Eq.~1 in the main text is $\epsilon_{0} = -nJ$, where $n=4$ is the number of nearest neighbours. For a field applied along the pseudo-easy-axis (the $[0\,0\,1]$ direction), a spin-flop transition from the ground state to a spin-canted state will occur at the field $\mu_{0}H_{\text{SF}}$. A schematic of the canted structure for spins at site $i$ and $j$ is shown in Fig.~\ref{fig: Flop_struct}. The average energy per spin in the spin-flop state for a classical vector model is,
\begin{dmath}
   \epsilon_{\text{SF}}= nJ \left\{  \sin\theta_{i}\sin\theta_{j}\left[\cos\phi_{i}\cos\phi_{j}+\sin\phi_{i}\sin\phi_{j}\right] + \cos\theta_{i}\cos\theta_{j}\right\} + D\left\{\sin^2\theta_{i}\sin^2\phi_{i} + \sin^2\theta_{j}\cos^2\phi_{j} \right\} - g\mu_{\text{B}}\mu_{0}H\left\{\cos\theta_{i} + \cos\theta_{j}\right\},
   \label{eq: E_SF}   
\end{dmath}
where $\left(\theta_{i,j},\phi_{i,j}\right)$ are $\mathbf{S}_{i,j}$ vectors expressed in spherical polar coordinates. $\theta_{i,j}$ are the angles between the spins $\mathbf{S}_{i,j}$ and the applied field. $\phi_{i,j}$ are the azimuthal angles which measure the rotation of the spins away from the $[1\,1\,0]$ about the direction of the applied field. \textcolor{black}{In conventional easy-axis systems, sites $i$ and $j$ are equivalent so $\theta_{i}=\theta_{j}$ and there is no ferromagnetic component of the spins perpendicular to the applied field, i.e. $\phi_{i} = \phi_{j} + \pi$. In our systems, the rotating SIA direction breaks the symmetry between the $i$ and $j$ sites. Spins will cant toward their respective easy-planes (away from the hard-axis $\hat{\mathbf{D}}_{i,j}$) leading to a ferromagnetic component perpendicular to the applied field.} However, for $D\ll nJ$, this ferromagnetic component will be small and we can approximate that $\phi_{i} \approx \phi_{j}+180 = \phi$ and $\theta_{i} \approx \theta_{j} = \theta$. The resulting expression for $\epsilon_{\text{SF}}$ is,
\begin{dmath}
   \epsilon_{\text{SF}}= nJ\cos^{2}\theta - nJ - D\cos^{2}\theta + D - 2g\mu_{\text{B}}\mu_{0}H\cos\theta.
   \label{eq: E_SF_aprrox}   
\end{dmath}
By differentiating $\epsilon_{\text{SF}}$ with respect to $\sin\theta$, Eq.~\ref{eq: E_SF_aprrox} is found to be minimised when $\cos\theta=2g\mu_{\text{B}}\mu_{0}H\cos/\left[ 2nJ-D\right]$. Magnetic saturation occurs when $\theta = 0$, therefore the saturation field is given by
\begin{dmath}
   \mu_{0}H_{1}^{\text{sat}} = \frac{2nJ -D}{g\mu_{B}}.
   \label{eq: H_z_sat}   
\end{dmath}
The spin-flop transition is expected to occur when $\epsilon_{\text{SF}} = \epsilon_{0}$. Substitution of the expression for $\cos\theta$ back in Eq.~\ref{eq: E_SF_aprrox} and equating to $\epsilon_{0}$ give the spin-flop field
\begin{dmath}
    \mu_{0}H_{\text{SF}} = \frac{\sqrt{2nJ D-D^{2}}}{g\mu_{B}}.
    \label{eq: Spin-flop}   
\end{dmath}
In reality, the spin-flop transition may deviate slightly from the prediction of Eq.~\ref{eq: Spin-flop}, as relaxing the approximations, $\phi_{i} \approx \phi_{j}+180 = \phi$ and $\theta_{i} \approx \theta_{j} = \theta$, may lead to a slight reduction in $\epsilon_{\text{SF}}$. As a result, the spin-flop transition may occur at a slightly lower field and the real value of $D$ will be slightly higher than calculated. However, Eq.~\ref{eq: Spin-flop} will be increasingly more accurate as $D/nJ\rightarrow0$.

\subsection{Monte-Carlo simulations}
\begin{figure}
    \centering
    \includegraphics[width= 0.72\linewidth]{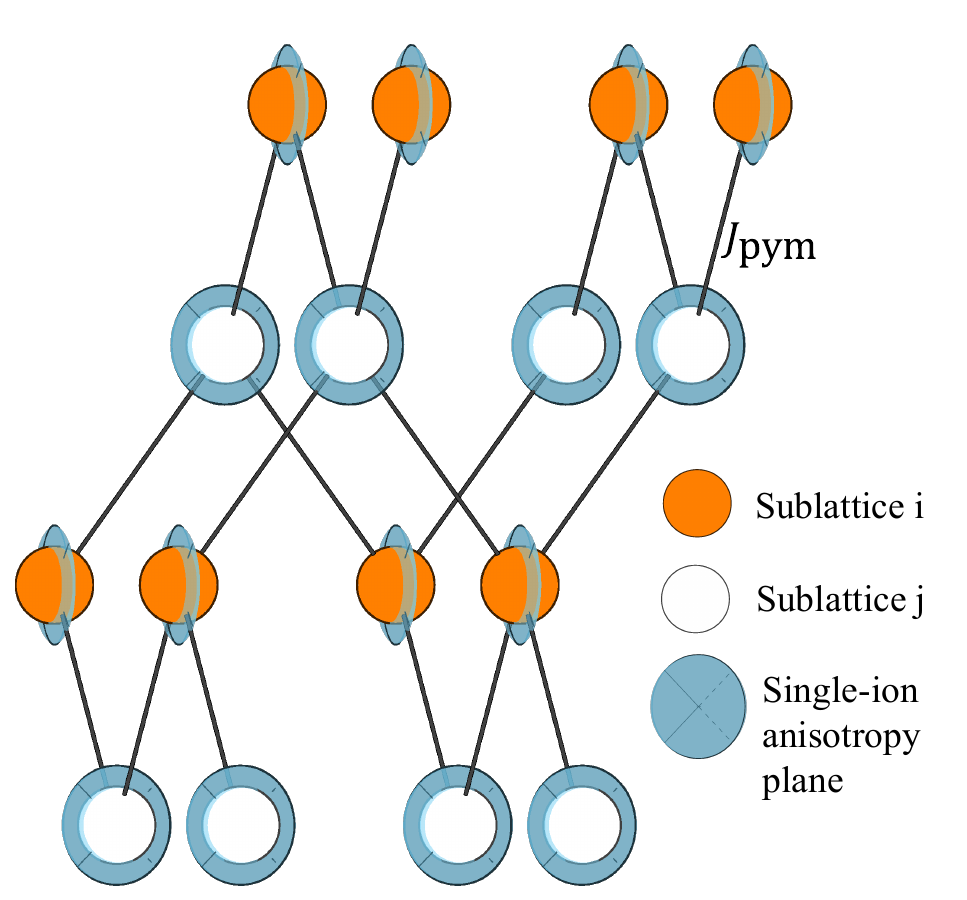}
    \caption{The 16-spin cluster used for classical Monte-Carlo simulation of the magnetisation in Ni$X_{2}$(pym)$_{2}$. Spins are arranged as two 8-spin sublattices and ions from one sublattice are coupled via Heisenberg exchange to the four nearest ions of the second sublattice. Blue circles represent the single-ion anisotropy plane which is rotated by $90^{\circ}$ between sublattices. The simulation uses a periodic boundary condition.}
    \label{fig: monte_struct}
\end{figure}

The powder-averaged magnetisation curves for Ni$X_{2}$(pym)$_{2}$ were computed using a classical Monte-Carlo routine written in Julia. The simulation utilises a cluster of 16 ions which are arranged as two sublattices of 8 ions as shown in Fig.~\ref{fig: monte_struct}. The magnetic moment of each ion is represented as classical unit vectors. The total energy of the cluster is calculated using the Hamiltonian in Eq.~1 with periodic boundary conditions and the simulation aims to find the spin configuration which minimises the energy of the whole cluster. Simulations were run iteratively, for $0\leq\mu_{0}|\mathbf{H}|\leq 60$\,T in $\Delta\mu_{0}|\mathbf{H}|=0.3$\,T steps.  

For each value of $|\mathbf{H}|$, the orientation of one spin is assigned a random direction. The change is accepted if it leads to a decrease in the cluster energy; otherwise, the change is only accepted with a probability given by $\text{exp}(-\Delta\epsilon/k_{\text{B}}T)$. Here $\Delta\epsilon$ is the change in energy and $T = 0.1$\,K. This process is iterated for every spin in the cluster and is repeated 4000 times. Finally, the longitudinal magnetisation $M$ of the cluster is calculated before increasing $|\mathbf{H}|$. 

For $0\leq \mu_{0}|\mathbf{H}| \leq 12$, a spin-flop transition is expected to occur. The collinear antiferromagnetic ground state in this region is a local energy minimum and the barrier to the global minimum is $> 0.1$\,K. Therefore, this routine fails to find the spin-flop state and simulated annealing is employed for these values of $\mu_{0}|\mathbf{H}|$. The powder average is calculated by averaging the magnetisation curves for all 100 different field orientations.
\bibliography{main.bib}